\documentclass[12pt, titlepage]{myarticle}

\usepackage{amsmath}
\usepackage{amsfonts}
\usepackage[pdftitle={A New Boundary Counterterm for Asymptotically AdS Spacetimes},pdfauthor={Robert A. McNees}]{hyperref}
\long\def\omit#1{}

\numberwithin{equation}{section}

\begin{document}

\preprint{Brown-HET-1460\\
%MCTP--xx-xx\\ 
{\tt hep-th/0512297}\\ }

\title{A New Boundary Counterterm for Asymptotically AdS Spacetimes}

\author{Robert McNees}
\oneaddress{Department of Physics,\\
Brown University,\\
Providence, RI 02912 \\ {~}\\
\email{mcnees@het.brown.edu}}

\Abstract{
We present a modified version of the boundary counterterm method for removing divergences from the action of an asymptotically $AdS$ spacetime. The standard approach renders the action finite but leaves diffeomorphism invariance partially broken if the dimension of the spacetime is odd. We show that this symmetry is restored by a new boundary counterterm, needed to cancel a divergence that appears in dimensional regularization. The result is a finite, diffeomorphism invariant action appropriate for gravitational physics. As an example we calculate the action for the Kerr-$AdS_5$ black hole. Unlike the standard boundary counterterm results, our action yields conserved charges that are consistent with the first law of black hole thermodynamics.
}

\maketitle

%\tableofcontents

%%%%%%%%%%%%%%%%%%%%%%%%%%%%%%%%%%%%%%%%%%%%%%%%%%%%%%%%%%%%%
%%%%%%%%%%%%%%%%%%%%%%%%%%%%%%%%%%%%%%%%%%%%%%%%%%%%%%%%%%%%%
\section{Introduction}
\label{sec:intro}
%%%%%%%%%%%%%%%%%%%%%%%%%%%%%%%%%%%%%%%%%%%%%%%%%%%%%%%%%%%%%
%%%%%%%%%%%%%%%%%%%%%%%%%%%%%%%%%%%%%%%%%%%%%%%%%%%%%%%%%%%%%

A common problem in Euclidean quantum gravity is the appearance of divergences in the gravitational action. For instance, the thermodynamics of black holes can be studied by approximating the black hole partition function as a saddle point of the Euclidean path integral \cite{Gibbons:1976ue}
\begin{equation}\label{BHrelation}
  \ZZ_{BH} \simeq \exp\left(-I_{BH}\right) ~.
\end{equation}
Unfortunately the right hand side of this relationship diverges, making its meaning unclear. This is usually addressed using a technique known as `background subtraction' \cite{Gibbons:1976ue}. First, one identifies a reference spacetime with the same asymptotics as the black hole spacetime. Then the action for the reference spacetime is used to normalize the relationship \eqref{BHrelation}
\begin{equation}\label{BHequality}
  \ZZ_{BH} = \exp\left(-\left(I_{BH}-I_{Ref}\right)\raisebox{12pt}{\,}\right) ~.
\end{equation}
Although both actions contain divergences, their finite difference gives a well defined approximation to the black hole partition function. Equation \eqref{BHequality} can be thought of as encoding the features of the black hole spacetime that were not already present in the reference spacetime.

A similar problem appears in applications of the $AdS$/CFT correspondence \cite{Maldacena:1997re,Gubser:1998bc,Witten:1998qj}. A common example of the correspondence is the duality between type IIB string theory with $AdS_5 \times S^5$ boundary conditions, and $\NN=4$ SYM in four dimensions with gauge group $SU(N)$. Because the duality maps strong coupling in one theory to weak coupling in the other, one may obtain the partition function for the SYM theory at strong `t Hooft coupling by evaluating the supergravity path integral
\begin{equation}\label{AdSCFTrelation}
  \ZZ_{SYM}\left[\dM,\gamma\right] \simeq \sum_{\MM}\,\int \DD g \exp\left(-I_{SUGRA}[\MM,g]\raisebox{12pt}{\,}\right) ~.
\end{equation}
The boundary conditions for the supergravity path integral are set by the spacetime $(\dM,\gamma)$ on which the field theory resides. In the limit of large $N$ and large `t Hooft coupling the right hand side of \eqref{AdSCFTrelation} is dominated by the saddle points of the tree-level supergravity action. However, as in the previous example, the supergravity action diverges, leaving the SYM partition function ill-defined.

Extracting a meaningful result from \eqref{AdSCFTrelation} requires a systematic method for regulating the supergravity action and removing its divergences, like the background subtraction used to normalize \eqref{BHrelation}. This is made easier by the fact that many features of the SYM theory are captured by pure gravity in five dimensions with a negative cosmological constant, rather than the full ten dimensional supergravity. In that case one usually employs an elegant technique known as the `boundary counterterm' method \cite{Henningson:1998gx,Balasubramanian:1999re,Emparan:1999pm,deHaro:2000xn}. This involves supplementing the gravity action that appears in \eqref{AdSCFTrelation} with a suitable counterterm
\begin{equation}
   \Gamma = I - I_{CT} ~.
\end{equation}
The counterterm, whose construction is reviewed in section \ref{sec:BoundaryCounterterms}, is a boundary term that cancels the divergences in the gravity action. Unlike the external reference spacetime used in background subtraction, the counterterm is intrinsic to the spacetime being studied. The SYM partition function is related to the new action $\Gamma$  by
\begin{equation}\label{AdSCFTequality}
  \ZZ_{SYM} = \exp\left(-\Gamma\right) ~.
\end{equation}
This result leads to several satisfying predictions of the $AdS$/CFT correspondence. For instance, it gives the correct conformal anomaly when the field theory resides on a spacetime with non-trivial curvature or topology.

The techniques described above represent two approaches to dealing with the problem of divergences in the gravitational action. Although they bear a superficial resemblance to one another they are qualitatively different. In background subtraction, for example, everything is defined relative to a reference spacetime. In certain cases the appropriate reference can be ambiguous, or unknown. The boundary counterterm method does not require a reference spacetime. On the other hand, the boundary counterterm method only applies to spacetimes with a non-zero cosmological constant. Although many authors have tried to extend the technique to asymptotically flat spacetimes \cite{Kraus:1999di}, it does not seem that this is possible in general\footnote{As this work was being completed reference \cite{Mann:2005yr} appeared, in which the authors propose a counterterm construction for asymptotically flat spacetimes.}. Background subtraction does not have this restriction; it works equally well for asymptotically flat spacetimes. Given these two different methods for addressing the same general problem, it is interesting to study cases for which both methods are applicable. If the two methods give different results it is important to understand why this is so, and whether the dissenting method can be amended to produce the correct physics.

A basis for comparing the two methods has been established by a number of authors who used the boundary counterterm method to study the conserved charges and thermodynamics of various asymptotically (locally) $AdS$ spacetimes \cite{Balasubramanian:1999re, Emparan:1999pm, Myers:1999ps, Kraus:1999di, Awad:1999xx, Awad:2000ac}. In cases where background subtraction can also be used, the two methods give nearly identical results. The one caveat involves asymptotically $AdS$ spacetimes with odd dimension\footnote{We will be almost exclusively concerned with asymptotically $AdS$ spacetimes whose dimension $d+1$ is odd. Their even-dimensional counterparts do not suffer from the various problems we discuss, for reasons that will be reviewed later in the paper. Unless we explicitly state otherwise, it should be assumed that the spacetime dimension is odd. }. Using the quasi-local stress tensor of Brown and York \cite{Brown:1992br}, the boundary counterterm method gives a mass for the spacetime that differs from the result one obtains using background subtraction. The difference between the two results is interpreted as the Casimir energy of the dual field theory on the boundary of the space \cite{Balasubramanian:1999re,Emparan:1999pm,Myers:1999ps}
\begin{equation}
   M_{BCT} = M_{BS} + E_{Casimir} ~.   
\end{equation}
The fact that this Casimir term is non-zero in the case of global $AdS$ has led some authors to question the validity of the boundary counterterm method \cite{Ashtekar:1999jx}. It is argued that a covariant procedure for determining the energy of an asymptotically $AdS$ spacetime cannot give a non-zero energy for global $AdS$. If this were the case it would imply the existence of a non-zero $d+1$-momentum that is invariant under all isometries of the global $AdS$ metric, and there is no such vector. More recently, the authors of \cite{Gibbons:2004ai} raised a new objection by pointing out that the 
Casimir term in the mass interferes with the usual thermodynamic interpretation of the Kerr-$AdS$ black hole. Specifically they showed that, while the conserved charges obtained using background subtraction satisfy the first law of black hole thermodynamics \cite{Bardeen:1973gs,Iyer:1994ys}
\begin{equation}
  dM_{BS} = T \, dS_{BS} + \sum_{i} \Omega_{i} \, d J^{i}_{BS} ~,
\end{equation}
the conserved charges obtained using the boundary counterterm method \cite{Awad:1999xx} do not.

The existence of a non-zero mass for global $AdS$ and the breakdown of the first law for the Kerr-$AdS$ solutions are the result of a subtlety in defining the correct counterterms for a gravitational theory. Whereas background subtraction gives an action that is invariant under diffeomorphisms of the spacetime, the boundary counterterm method does not. The boundary counterterm method is engineered to give an action that is invariant under transformations on the boundary that are symmetries of the dual field theory. A generic diffeomorphism of the spacetime induces a transformation on the boundary which is not a symmetry of the dual field theory, and therefore not a symmetry of $\Gamma$.

In this paper we show that the boundary counterterm method can be used to construct a gravitational action that is both finite and diffeomorphism invariant. Not surprisingly, this action agrees with the action obtained from background subtraction when the two can be compared. Furthermore, the quasi-local stress tensor obtained from this action gives conserved charges that agree with a number of other approaches \cite{Das:2000cu,Deser:2005jf,Deruelle:2004mv,Hollands:2005wt,Hollands:2005ya} and satisfy the first law of black hole thermodynamics. We obtain the action by including a new counterterm that is needed to cancel a divergence that appears in dimensional regularization. When the dimension is continued back to the physically relevant value this contribution to the counterterm action is finite. It is tempting to think of this finite term as representing a choice of renormalization scheme, along the lines of the ambiguity in the boundary counterterm construction first pointed out in \cite{Balasubramanian:1999re}. In fact, this is not quite the case, as we will explain later in the paper.
Finally, our modification does not alter the intrinsic character of the counterterm construction so that, unlike background subtraction, it does not require the use of a reference spacetime.

We proceed as follows. In section \ref{sec:Divergences} we review the properties of asymptotically $AdS$ spacetimes. We also explain the problems that arise when one tries to evaluate the `usual' gravitational action for these spacetimes.  In section \ref{sec:BoundaryCounterterms} we summarize the boundary counterterm method and then explain its derivation from a solution of the Hamilton-Jacobi equation. We also show that the standard choice of counterterms, which is a minimal subtraction, leads to an action which is not invariant under certain diffeomorphisms. The restoration of this symmetry, which requires a new counterterm, is described in section \ref{sec:CountertermHamiltonian}. In section \ref{sec:KerrAdS5} we use our modified action to study the Kerr-AdS black hole and obtain, without modification, the first law of thermodynamics. Finally, we discuss some implications of our results in section \ref{sec:Conclusion}. Some technical results and background material are contained in two appendices.

The first two sections of this paper contain a large amount of background material. Readers who are already familiar with boundary counterterms and the properties of asymptotically $AdS$ spacetimes can review the conventions and definitions established in section \ref{sec:Conventions}, and then proceed directly to section \ref{sec:CountertermHamiltonian}.

%%%%%%%%%%%%%%%%%%%%%%%%%%%%%%%%%%%%%%%%%%%%%%%%%%%%%%%%%%%%%
%%%%%%%%%%%%%%%%%%%%%%%%%%%%%%%%%%%%%%%%%%%%%%%%%%%%%%%%%%%%%
\section{Asymptotically Anti-de Sitter Spacetimes}
\label{sec:Divergences}
%%%%%%%%%%%%%%%%%%%%%%%%%%%%%%%%%%%%%%%%%%%%%%%%%%%%%%%%%%%%%
%%%%%%%%%%%%%%%%%%%%%%%%%%%%%%%%%%%%%%%%%%%%%%%%%%%%%%%%%%%%%

In this paper we are interested in spacetimes $(\MM,g)$ that are asymptotically Anti-de Sitter. Our goal is to show that they possess a well defined action that is both finite and diffeomorphism invariant. The construction of this action will depend crucially on certain properties of these spacetimes, which we must review before we can proceed. A useful (but incomplete) list of references to the extensive literature on this subject includes \cite{Ellis:1973xx, Ashtekar:1984aa, Magnon:1985sc, Henneaux:1985tv, Ashtekar:1999jx, Henningson:1998gx, Graham:1999pm, Bengtsson:1998xx, Anderson:2004wj, Anderson:2004yi, Papadimitriou:2005ii}.

We begin with a brief review of Anti-de Sitter space ($AdS$), followed by a precise definition of asymptotically $AdS$ spacetimes. Next we review asymptotic coordinate systems for these spacetimes and introduce our conventions for foliations in the vicinity of spatial infinity. Finally, we discuss some of the inconsistencies that arise when one tries to evaluate the standard form of the gravitational action for an asymptotically $AdS$ spacetime.

%%%%%%%%%%%%%%%%%%%%%%%%%%%%%%%%%%%%%%%%%%%%%%%%%%%%%%%%%%%%%
\subsection{Properties Of Asymptotically \texorpdfstring{$AdS$}{AdS} Spacetimes}
\label{sec:Conventions}
%%%%%%%%%%%%%%%%%%%%%%%%%%%%%%%%%%%%%%%%%%%%%%%%%%%%%%%%%%%%%

The best way to set the stage for a discussion of asymptotically $AdS$ spacetimes is to first review the properties of $AdS$ itself. The starting point is pure gravity in $d+1$ dimensions with a negative cosmological constant. The Einstein equations are
\begin{equation} 
   R_{\mu\nu} - \frac{1}{2}\,g_{\mu\nu}\,\left(R - 2 \Lambda\right) = 0 ~.
\end{equation}
We will follow the conventional normalization and express the cosmological constant in terms of a length scale $\ell$ by
\begin{equation}\label{CosmologicalConstant}
  \Lambda = - \frac{d(d-1)}{2 \,\ell^2}
\end{equation}
in which case Einstein's equation reduces to
\begin{equation}\label{Einstein}
  R_{\mu\nu} = - \frac{d}{\,\ell^2} \, g_{\mu\nu} ~.
\end{equation}
The simplest solution of this equation is Anti-de Sitter space. It is the maximally symmetric space of constant negative curvature, with Riemann tensor given by
\begin{equation}\label{AdSRiemann}
   R^{\lambda}_{\,\,\,\mu\sigma\nu} = - \frac{1}{\ell^2}\,\left(g_{\mu\nu} \, \delta^{\lambda}_{\,\,\,\sigma} 
      - g_{\mu\sigma} \, \delta^{\lambda}_{\,\,\,\nu} \right) ~.
\end{equation}
A useful parameterization of $AdS$ is given by the global coordinates\footnote{Technically, this represents a universal cover of $AdS$, obtained by `unwinding' a periodic time coordinate and replacing it with the non-compact time coordinate $t$. When we refer to $AdS$ we always refer to the causal spacetime obtained in this manner.}
 $(r,t,\sigma_{d-1})$, where $0 \leq r < \infty$ and $-\infty < t < \infty$, while $\sigma_{d-1}$ represents coordinates on a sphere $S^{d-1}$ of radius $\ell$. 
In these coordinates the metric on $AdS$ is
\begin{equation}\label{AdSmetric}
  ds^2 = \left(1 + \frac{r^2}{\ell^2}\right)^{-1} \nts dr^2 - \left(1 + \frac{r^2}{\ell^2}\right) dt^2 
   + \frac{r^2}{\ell^2} \, d \sigma_{d-1}^{\,2} ~.
\end{equation}
There is a second-order pole in the metric at $r \to \infty$, where $AdS$ possesses a conformal boundary with topology $\BR \times S^{d-1}$. This divergence in the metric at spatial infinity is common to all solutions of \eqref{Einstein} and, as described below, implies that the metric \eqref{AdSmetric} does not induce a unique metric on the conformal boundary.

With this description of $AdS$ in mind we want to establish a definition for an asymptotically $AdS$ spacetime \cite{Ashtekar:1984aa,Magnon:1985sc,Henneaux:1985tv,Ashtekar:1999jx,Graham:1999jg}. Following the discussion in \cite{Ashtekar:1999jx} we say that a spacetime $(\MM,g)$ that satisfies \eqref{Einstein} is asymptotically $AdS$ if there exists a spacetime $(\hMM,\hg)$, where $\hMM$ is a manifold with boundary $\dhM$, such that:
\begin{enumerate}
 \item The manifold $\MM$ can be identified with the interior of $\hMM$ by means of a diffeomorphism from $\MM$ onto $\hMM - \dhM$.
 \item There exists a positive function $\Omega$ on $\hMM$, known as a `defining function', such that $\hg_{\mu\nu} = \Omega^2 \, g_{\mu\nu}$. Furthermore, $\Omega$ has a first-order zero and a non-vanishing gradient on $\dhM$.
 \item The Weyl tensor $\hat{C}_{\lambda\mu\sigma\nu}$ on $\hMM$, constructed from $\hg_{\mu\nu}$, is such that $\Omega^{3-d}\hat{C}_{\lambda\mu\sigma\nu}$ is smooth on $\hMM$, and $\hat{C}_{\lambda\mu\sigma\nu}$ vanishes on $\dhM$.
 \item The boundary $\dhM$ is topologically $\BR \times S^{d-1}$.
\end{enumerate}
We will distinguish between $(\MM,g)$ and $(\hMM,\hg)$ by referring to them as the `physical' and `unphysical' spacetimes, respectively.

The primary benefit of this definition is that it identifies $\MM$ with the interior of $\hMM$, which allows one to work with the unphysical spacetime $(\hMM,\hg)$. This neatly circumvents many technical difficulties that arise when working with the physical spacetime. For instance, the conformal boundary of $\MM$, which we denote by $\dM$, is located at spatial infinity and is not part of $\MM$. But the boundary $\dhM$ of the unphysical spacetime is defined as a finite value of a spacelike coordinate: $\Omega=0$. Thus, working with the unphysical spacetime allows one to avoid some of the infinite limits associated with attaching the surface $\dM$ to $\MM$. If we insist on working with the physical spacetime, we find that we cannot extend the physical metric to $\dM$ because it diverges at spatial infinity. The defining function, whose properties are established in the second part of the definition, addresses this problem by insuring that the unphysical metric $\hg_{\mu\nu}$ is finite and well-defined on all of $\hMM$, including $\dhM$. Having established the framework of the unphysical spacetime, the third point in the definition is simply the requirement that the Riemann tensor for an asymptotically $AdS$ spacetime should approach the maximally symmetric form \eqref{AdSRiemann} at spatial infinity. On the other hand, the Riemann tensor will generically have a non-vanishing Weyl tensor on the interior of the space and take the form
\begin{equation} \label{Riemann}
  R^{\lambda}_{\,\,\mu\sigma\nu} = C^{\lambda}_{\,\,\mu\sigma\nu} - \frac{1}{\ell^2}\,\left(g_{\mu\nu} \, 
    \delta^{\lambda}_{\,\,\sigma} - g_{\mu\sigma} \, \delta^{\lambda}_{\,\,\nu} \right) ~.
\end{equation}
The first three parts of the definition already imply that $(\MM,g)$ is locally asymptotic to $AdS$. To complete the definition we must include the fourth point and require that the topology of the boundary is $\BR \times S^{d-1}$. This insures that the spacetime is asymptotically $AdS$ in the global sense~\footnote{The phrase `asymptotically $AdS$' is often used in the literature to describe spacetimes which are only asymptotically locally $AdS$. For a discussion of this more general class of spacetimes see \cite{Papadimitriou:2005ii}, and references therein.}. This also implies that the conformal boundary of an asymptotically $AdS$ spacetime has vanishing Euler number. This will play an important role later in the paper.

There is a natural ambiguity in the construction of the unphysical spacetime that reflects a real and important property of asymptotically $AdS$ spacetimes. As stated above, all metrics that solve \eqref{Einstein} exhibit a second-order pole at spatial infinity. Thus, even if we attach the conformal boundary $\dM$ to $\MM$ the metric is not defined there. To address this we introduce a defining function that cancels the divergence and gives a finite but unphysical metric $\hg$. But the properties of $\Omega$ are not sufficient to determine it uniquely. In other words, there are many possible choices of unphysical spacetime, none of which are preferred over the others. The equivalent statement in the physical spacetime is that the metric on an asymptotically $AdS$ spacetime does not induce a unique metric on the conformal boundary. Rather, it endows $\dM$ with a conformal class of metrics. To see this, consider a defining function $\Omega$ and the corresponding unphysical metric $\hg_{\mu\nu} = \Omega^2 g_{\mu\nu}$ on $\hMM$. The pullback of this metric onto the surface $\Omega=0$ induces a metric $\hat{h}_{\mu\nu}$ on $\dhM$. But the choice of defining function $\Omega$ is not unique. Any non-vanishing function $e^{\omega}$ on $\hMM$ can be used to obtain a new defining function $\tilde{\Omega} = e^{\omega}\,\Omega$. The unphysical metric associated with this new defining function is related to $\hg_{\mu\nu}$ by a conformal rescaling
\begin{equation}
   \tg_{\mu\nu} = e^{2 \omega} \hg_{\mu\nu}
\end{equation}
The pullback of $\tg_{\mu\nu}$ also induces a metric on $\dhM$, related to $\hat{h}_{\mu\nu}$ by
\begin{equation}
   \tilde{h}_{\mu\nu} = \left.e^{2 \omega}\right|_{\dhM} \hat{h}_{\mu\nu}
\end{equation}
Therefore a particular choice of defining function merely identifies a representative of a conformal class of metrics related by (non-singular) local Weyl rescalings. As was shown in \cite{Ashtekar:1999jx}, there is always a choice of defining function for which the induced metric on $\dhM$ is manifestly conformally flat. Since the Weyl tensor is invariant under local rescalings of the metric, this result applies to the entire conformal class of metrics on the boundary. Thus, the metric on an asymptotically $AdS$ spacetime induces a conformal class of metrics on the boundary, and that class of metrics is conformally flat.

All of the properties of asymptotically $AdS$ spacetimes mentioned in this section will be used, either explicitly or implicitly, throughout this paper. But two points will play an important role and should be re-emphasized before proceeding:
\begin{enumerate}
  \item The topology of the conformal boundary implies that its Euler number is zero.
  \item Any member of the conformal class of metrics induced on the conformal boundary has a vanishing Weyl tensor. 
\end{enumerate}
Both of these follow from the definition of an asymptotically $AdS$ spacetime. The first point is a simple consequence of the topology, and the second can be shown in a variety of ways using nothing more than local differential geometry \cite{Ashtekar:1999jx,Skenderis:1999nb}. As a final note, in later sections of this paper we will often consider the Euclidean section of an asymptotically $AdS$ spacetime, where the non-compact time coordinate $t$ is replaced with a periodic imaginary time coordinate $\tau$. This changes the factor of $\BR$ in the boundary topology to an $S^1$, but it does not change any of the conclusions of this section.

%%%%%%%%%%%%%%%%%%%%%%%%%%%%%%%%%%%%%%%%%%%%%%%%%%%%%%%%%%%%%
\subsection{Asymptotic Coordinate Systems}
\label{sec:Foliations}
%%%%%%%%%%%%%%%%%%%%%%%%%%%%%%%%%%%%%%%%%%%%%%%%%%%%%%%%%%%%%

In the previous section we gave a definition of asymptotically $AdS$ spacetimes that uses an unphysical spacetime
to establish many of their properties. Despite the utility of the unphysical spacetime, some of the calculations that appear later in this paper are better suited to the physical spacetime. In this section we will establish a useful coordinate system to perform these calculations in.

The defining function $\Omega$ that relates the physical and unphysical metrics has a non-vanishing gradient at the boundary $\dhM$. This allows us to construct an asymptotic coordinate system, with $\Omega$ as a coordinate normal to the boundary, in which the boundary is located at $\Omega=0$. Thus, near the boundary we can always choose coordinates in which the metric on $\MM$ takes the Fefferman-Graham form \cite{Graham:1985CI}
\begin{equation}\label{FGmetric}
  ds^2 = \frac{\ell^2}{4 \rho^2}\, d \rho^2 + \frac{\ell}{\rho}\,H_{ab}(x,\rho)\,dx^a dx^b ~.
\end{equation}
However, for the spacetimes we consider in this paper it is convenient to use a coordinate system in which the metric resembles the $r \gg \ell$ limit of \eqref{AdSmetric}
\begin{equation}\label{CoordSystem}
  ds^2 = \frac{\ell^2}{r^2}\,h_{rr}(r)\,dr^2 + \frac{r^2}{\ell^2}\,h_{ab}(x,r)\,dx^a\,dx^b
\end{equation}
In this coordinate system the boundary is located at $r \to \infty$, where the metric exhibits the characteristic second-order pole. The functions $h_{ab}$ are smooth and yield a $d$-dimensional metric, conformal to the Einstein static universe, in the $r \to \infty$ limit. Although we could perform a coordinate redefinition that sets the function $h_{rr}=1$, we will leave it arbitrary for now and simply note that in the $r \to \infty$ limit it approaches unity. 
From now on we will adopt our index notation to this coordinate system, using Greek letters $\mu,\nu,\ldots$ to denote $d+1$ dimensional indices and Latin letters $a,b,\ldots$ for $d$-dimensional indices.

The coordinate system \eqref{CoordSystem} provides a useful foliation of $\MM$ in the region $r \gg \ell$.
The leaves of this foliation are timelike hypersurfaces of constant $r$ that we denote $\Sigma_r$. 
The embedding of $\Sigma_r$ in $\MM$ is described locally by a spacelike unit normal vector $n^\mu$ that takes the form
\begin{equation}
  n^\mu = \frac{r}{\ell \, \sqrt{h_{rr}}} \, \delta_{\mu r} ~.
\end{equation}
The pullback of the physical metric $g_{\mu\nu}$ induces a metric $\gamma_{\mu\nu}$ on $\Sigma_r$ given by
\begin{equation}
  \gamma_{\mu\nu} = g_{\mu\nu} - n_{\mu}\,n_{\nu} ~.
\end{equation}
In the coordinate system \eqref{CoordSystem} this is simply
\begin{equation}\label{InducedMetric}
  \gamma_{ab}(x,r) = \frac{r^2}{\ell^2}\,h_{ab}(x,r) ~.
\end{equation}
The intrinsic curvature tensors constructed from $\gamma_{ab}$ and its derivatives along $\Sigma_r$ are denoted by $\RR^{d}_{\,\,cab}$, $\RR_{ab}$, and $\RR$, while the $d$-dimensional covariant derivative, compatible with $\gamma_{ab}$, is $\DD_a$. Our convention for the extrinsic curvature associated with the embedding of $\Sigma_r$ in $\MM$ is
\begin{equation}\label{ExtrinsicCurvature}
  K_{\mu\nu} = \nabla_{\mu} n_{\nu} - n_{\mu} a_{\nu}
\end{equation}
where $a_\nu = n^{\lambda}\nabla_{\lambda}n_{\nu}$. Expressed in terms of the normal derivative of $\gamma_{ab}$ this is just
\begin{equation}\label{ExtrinsicCurvature2}
   K_{ab} = \frac{1}{2}\,n^{\mu} \partial_{\mu} \gamma_{ab} ~.
\end{equation}
The trace of the extrinsic curvature, which appears in the gravitational action, is given by $K = \nabla_{\mu} n^{\mu}$.

By construction, the conformal boundary of the asymptotically $AdS$ spacetime corresponds to the $r \to \infty$ limit of the surface $\Sigma_r$. Given a choice of defining function $\Omega$, the induced metric on $\Sigma_r$ singles out a representative $\hat{h}_{ab}(x)$ of the conformal class of metrics on $\dM$. For instance, in the coordinate system \eqref{CoordSystem} one may choose $\Omega = \ell / r$, in which case $\hat{h}_{ab}(x)$ is given by
\begin{equation}\label{InducedRepRelation}
  \hat{h}_{ab}(x) = \lim_{r \to \infty} \frac{\ell^2}{r^2}\,\gamma_{ab}(x,r) = \lim_{r \to \infty} h_{ab}(x,r) ~.
\end{equation}
The definition of an asymptotically $AdS$ spacetime guarantees that the boundary Weyl tensor constructed from this metric vanishes. This does \emph{not} imply that the intrinsic Weyl tensor on $\Sigma_r$ vanishes, though it will be proportional to terms that vanish in the $r \to \infty$ limit.

%%%%%%%%%%%%%%%%%%%%%%%%%%%%%%%%%%%%%%%%%%%%%%%%%%%%%%%%%%%%%
\subsection{The Gravitational Action} 
\label{sec:OnShell}
%%%%%%%%%%%%%%%%%%%%%%%%%%%%%%%%%%%%%%%%%%%%%%%%%%%%%%%%%%%%%

The starting point of section \ref{sec:Conventions} was the Einstein equations \eqref{Einstein}. We usually assume that these equations have been obtained by extremizing a gravitational action of the form
\begin{equation}\label{GravityAction}
I = -\frac{1}{2 \kappa^2}\,\int_{\MM} \nts \nts d^{\,d+1}x \sqrt{g} \left(R - 2 \Lambda\right) - 
\frac{1}{\kappa^2} \, \int_{\dM} \bns d^{d}x \sqrt{\gamma}\,K ~.
\end{equation}
This action comprises two terms: the Einstein-Hilbert term in the bulk, and the Gibbons-Hawking-York term \cite{York:1972sj,Gibbons:1976ue} on the boundary. In this expression $\gamma_{\mu\nu}$ is the $d$-dimensional metric induced on the boundary by $g_{\mu\nu}$, the $d+1$-dimensional metric on $\MM$. Under a small variation of the metric $g_{\mu\nu} \to g_{\mu\nu} + \delta g_{\mu\nu}$ the change in the action is
\begin{equation}\label{ActionVariation}
 \delta I = \frac{1}{2\,\kappa^2}\,\int_{\MM} \nts \nts d^{d+1}x \sqrt{g}\left(R^{\mu\nu} - \frac{1}{2}
       \,g^{\mu\nu}\left(R-2 \Lambda\right)\right) \, \delta g_{\mu\nu}
       + \frac{1}{2\,\kappa^2}\,\int_{\dM} \bns d^{d}x \sqrt{\gamma} \left(K^{\mu\nu} - \gamma^{\mu\nu} K \right) \, 
        \delta \gamma_{\mu\nu}
\end{equation}
The role of the GHY term is to insure that there are no boundary terms in \eqref{ActionVariation} that are proportional to the normal derivative of $\delta \gamma_{\mu\nu}$. Thus, the action \eqref{GravityAction} appears to be extremized by solutions of \eqref{Einstein} with Dirichlet boundary conditions on the metric at $\dM$.

Upon closer examination the conclusions of the previous paragraph cannot be entirely correct. The definition of the action
and the description of the variational principle rely on assumptions that are not consistent with the solutions of the resulting equation of motion. There are three main causes for concern \cite{Henningson:1998gx,Papadimitriou:2005ii}. First, the GHY term is a functional of the induced metric on $\dM$. But metrics that solve \eqref{Einstein} diverge at the boundary and fail to induce a unique metric there. Second, since solutions of \eqref{Einstein} induce a conformal class of metrics on $\dM$, the imposition of Dirichlet boundary conditions on the metric seems overly restrictive. A more suitable action might allow for boundary conditions which admit variations within a fixed conformal class \cite{Papadimitriou:2005ii}. Finally, the bulk term in \eqref{GravityAction} diverges when evaluated for solutions of \eqref{Einstein}. Because the integrand is constant the action diverges in the infra-red due to the infinite volume of $\MM$.

As was recently shown in \cite{Papadimitriou:2005ii} these problems are all very closely related. The boundary counterterm technique used to address the divergences of the action also results in a consistent variational principle, but with modified boundary conditions. Actions constructed in this way are designed to respect the symmetries of the dual field theory and not the full diffeomorphism invariance of the gravitational theory. Understanding the physics of asymptotically $AdS$ spacetimes requires an action principle consistent with solutions of \eqref{Einstein}, like the actions presented in \cite{Balasubramanian:1999re,Emparan:1999pm,Papadimitriou:2005ii}, but also invariant under diffeomorphisms. In this paper we will construct an action with these properties.

Our approach begins by addressing the divergences in the action. To do this we must introduce a regulator that allows us to work with finite quantities. In the coordinate system \eqref{CoordSystem} this is achieved by putting a cut-off on the coordinate $r$ at some large but finite value
\begin{equation}
   r \leq r_c  \hspace{.25cm} \textrm{with} \hspace{.25cm} r_c \gg \ell ~.
\end{equation}
The result is a regulated spacetime $\MM_c$ whose boundary is the hypersurface $\Sigma_{c}$ defined by the cut-off. The action \eqref{GravityAction} can then be evaluated on the regulated spacetime, with the GHY term evaluated on $\Sigma_{c}$. Direct calculation gives a $d$-dimensional integral whose integrand contains a finite number of terms proportional to positive powers of the cut-off
\begin{equation}
\label{divergences}
  I = \frac{1}{\kappa^2} \, \int_{\Sigma_c} \nts \nts d^{d}x\,\left( c_0 \, r_c^{\,d} + c_2 \, r_c^{\,d-2} + c_4 \, 
       r_c^{\,d-4} +
\ldots \right) ~.
\end{equation}
Specifically, the regulated action generically contains $[(d+1)/2]$ terms which scale as positive powers\footnote{The action for an asymptotically locally $AdS$ spacetime, or the action for gravity plus certain forms of matter on an asymptotically $AdS$ spacetime, may also contain logarithmic divergences that are not distinguished by the power counting argument given here.} of $r_c$, while the `$\ldots$' represent terms which are finite or zero when $r_{c} \to \infty$. The structure of these terms is very suggestive when one considers the scaling of the induced metric on $\Sigma_c$. Since the cut-off satisfies $r_c \gg \ell$, the induced metric scales as $\gamma_{ab} \sim r_c^{\, 2}$. Similarly, the inverse metric $\gamma^{ab}$ scales as $r_c^{\,-2}$, and the factor of $\sqrt{\gamma}$ appearing in the covariant volume element scales as $r_c^{\,d}$. Thus, the divergent terms in \eqref{divergences} resemble an expansion in factors of the inverse metric on $\Sigma_c$. This observation motivates the boundary counterterm approach to removing these divergent terms from the action, which we will describe in section \ref{sec:BoundaryCounterterms}.

%%%%%%%%%%%%%%%%%%%%%%%%%%%%%%%%%%%%%%%%%%%%%%%%%%%%%%%%%%%%%
\subsection{A Note On Cut-Off Dependence} 
\label{sec:CutOff}
%%%%%%%%%%%%%%%%%%%%%%%%%%%%%%%%%%%%%%%%%%%%%%%%%%%%%%%%%%%%%

In the rest of the paper we will always employ the regulator described in the previous section. Thus, $\MM$ will be replaced by the regulated spacetime $\MM_c$ whose boundary is $\Sigma_c$. The $r_c \to \infty$ limit, which removes the cut-off and recovers the full spacetime, will be taken at the end of calculations after all cut-off dependent terms have been addressed. These calculations routinely involve asymptotic expansions, either in powers of the cut-off or in the inverse metric. We will always drop terms in these expansions which vanish as the cut-off is removed. Unless explicitly stated otherwise we do not distinguish between quantities that vanish identically, to all orders in the cut-off, and quantities that vanish in the $r_c \to \infty$ limit.

%%%%%%%%%%%%%%%%%%%%%%%%%%%%%%%%%%%%%%%%%%%%%%%%%%%%%%%%%%%%%
%%%%%%%%%%%%%%%%%%%%%%%%%%%%%%%%%%%%%%%%%%%%%%%%%%%%%%%%%%%%%
\section{Boundary Counterterms}
\label{sec:BoundaryCounterterms}
%%%%%%%%%%%%%%%%%%%%%%%%%%%%%%%%%%%%%%%%%%%%%%%%%%%%%%%%%%%%%
%%%%%%%%%%%%%%%%%%%%%%%%%%%%%%%%%%%%%%%%%%%%%%%%%%%%%%%%%%%%%

In this section we review two approaches to the boundary counterterm method for removing divergences from the on-shell action. Both constructions lead to a minimal subtraction scheme, which fails to preserve diffeomorphism invariance of the gravitational theory.

%%%%%%%%%%%%%%%%%%%%%%%%%%%%%%%%%%%%%%%%%%%%%%%%%%%%%%%%%%%%%
\subsection{Boundary Terms as Counterterms}
\label{sec:BoundaryTermsasCounterterms}
%%%%%%%%%%%%%%%%%%%%%%%%%%%%%%%%%%%%%%%%%%%%%%%%%%%%%%%%%%%%%

In \cite{Balasubramanian:1999re,Emparan:1999pm,Henningson:1998gx} a new technique was proposed for removing the divergences
from the on-shell action. By supplementing the gravity action \eqref{GravityAction} with a set of boundary counterterms, one obtains a renormalized action $\Gamma$ that is finite on-shell
\begin{equation}\label{GeneralRenormalizedAction}
    \Gamma = I - I_{CT} ~.
\end{equation}
The `counterterm action' is a boundary integral whose integrand is a local function of the induced metric
$\gamma_{ab}$ and its boundary derivatives
\begin{equation}\label{intrinsicfunctional}
   I_{CT} = \frac{1}{\kappa^2} \, \int_{\Sigma_c} \nts \nts d^{d} x \, \sqrt{\gamma} \, \LL_{CT}(\gamma, \partial_{a} \gamma) ~.
\end{equation}
The asymptotic behavior \eqref{InducedMetric} of $\gamma_{ab}$ gives the counterterm action the correct cut-off dependence to cancel the power-law divergent terms in the on-shell action \eqref{divergences}.

The general functional form of the counterterm action can be determined by expanding the Lagrangian density $\LL_{CT}$ in powers of the inverse metric
\begin{equation}
\label{1stCTexpansion}
   I_{CT} = \frac{1}{\kappa^2} \, \int_{\Sigma_c} \nts \nts d^{d}x \, \sqrt{\gamma} \left(\LL_{CT}^{(0)} + \LL_{CT}^{(2)} +
\LL_{CT}^{(4)} + \ldots \right) ~.
\end{equation}
A term in this expansion with superscript $(2n)$ contains $n$ factors of the inverse metric. 
A useful basis for this expansion consists of polynomials built from the intrinsic curvature tensors and their derivatives, with constant coefficients. The first few terms are
\begin{eqnarray}
\label{LCTbasis}
    \LL_{CT}^{(0)} & = & c_{0} \\ \nonumber
    \LL_{CT}^{(2)} & = & c_{1} \, \RR \\ \nonumber
    \LL_{CT}^{(4)} & = & c_{2}\,\RR^{2} + c_{3} \, \RR^{ab}\RR_{ab} + c_{4}\, \RR^{abcd} \RR_{abcd}  ~.
\end{eqnarray}
Using the arguments in section \ref{sec:OnShell} these terms scale with the cut-off as
\begin{equation}
     \sqrt{\gamma}\,\LL_{CT}^{(2n)} \sim r_{c}^{d-2n}
\end{equation}
so that terms with $2n > d$ vanish as the cut-off is removed. The counterterm action then comprises a finite number of local functionals of the boundary metric that are manifestly invariant under diffeomorphisms of the boundary coordinates.

Determining the coefficients $c_i$ that appear in \eqref{LCTbasis} gives the exact form of the counterterm action. One way to accomplish this is to explicitly cancel the divergences that appear in the on-shell action for a sufficiently general class of asymptotically $AdS$ metrics \cite{Balasubramanian:1999re, Emparan:1999pm, Liu:1998bu, Hyun:1998vg}. The resulting expression for the counterterm action is
\begin{eqnarray}\label{ctaction}
  I_{CT} & = & \frac{1}{\kappa^2} \, \int_{\Sigma_c} \nts \nts d^{d}x \, \sqrt{\gamma} \left(\frac{d-1}{\ell} + 
               \frac{\ell}{2(d-2)} \,\RR \right. \\ \nonumber
             &     &        \hspace{2cm}\left. + \frac{\ell^3}{2(d-2)^2 (d-4)}\left( \RR^{ab} \RR_{ab} -
                   \frac{d}{4(d-1)}\,\RR^2\right)  + \ldots \right) ~.
\end{eqnarray}
where `$\ldots$' denotes higher order terms in the inverse metric expansion. These higher order terms are easily
determined using a simple iterative procedure given in \cite{Kraus:1999di}. Note that the dimension of the boundary has been left arbitrary in \eqref{ctaction}. As stated above, a finite number of these terms are proportional to positive powers of the cut-off for a given $d$. Keeping only these terms defines a \emph{minimal subtraction} that removes the divergences \cite{Balasubramanian:1999re,Emparan:1999pm} from the action for any asymptotically (locally) $AdS_{d+1}$ spacetime
\begin{equation}\label{MSRenAction}
    \Gamma_{MS} = I - I_{MS} ~.
\end{equation}
This is the standard prescription for boundary counterterms in applications of the AdS/CFT correspondence.
The counterterm action in the minimal subtraction scheme is given by
\begin{equation}\label{MSCTaction}
   I_{MS} = \frac{1}{\kappa^2} \, \int_{\Sigma_c} \nts \nts d^{d}x \, \sqrt{\gamma} 
      \sum_{n=0}^{\left[\frac{1}{2}(d-1)\right]}\LL_{CT}^{(2n)} ~.
\end{equation}
The upper limit on the sum is the integer part of $\frac{1}{2}(d-1)$. Thus, in $d+1=3$ bulk dimensions only the first term in \eqref{ctaction}, proportional to the boundary volume element, is required. In four and five dimensions the counterterm involving the Ricci scalar of the boundary metric is also needed, while all of the terms shown in \eqref{ctaction} are required in six and seven dimensions. 

There is an important subtlety that applies when the spacetime dimension is odd, $d+1=2k+1$. In that case the dimension of the boundary is even, $d=2k$. Power counting then implies that terms in the counterterm action containing $k$ factors of the inverse metric are independent of the cut-off
\begin{equation}\label{AmbiguousTerm}
      \sqrt{\gamma}\,\LL_{CT}^{(2k)} \sim r_{c}^0 ~.
\end{equation}
The authors of \cite{Balasubramanian:1999re} point out that this represents the usual ambiguity associated with the choice of renormalization scheme. Minimal subtraction, which does not include a finite term like \eqref{AmbiguousTerm}, is one such choice. But one is always free to add or remove finite terms from the divergent counterterms to impose a different renormalization scheme. This is more apparent when we rearrange \eqref{GeneralRenormalizedAction} to read
\begin{equation}\label{NewActionExpression}
   I = \Gamma + I_{CT} ~.
\end{equation}
A choice of renormalization scheme is simply some criteria for whether we associate a particular finite term in the regulated action with $\Gamma$ or with $I_{CT}$.

The analysis in this section can be thought of as an attempt to determine the on-shell action's dependence on local functionals of the boundary metric, whose asymptotic behavior is assumed to be the source of the cut-off dependent terms in \eqref{divergences}. But, as discussed in the introduction, we are concerned with producing both a finite \emph{and} diffeomorphism invariant action. Because the on-shell action should be diffeomorphism invariant, its functional dependence on the diffeomorphism covariant boundary metric is heavily constrained. In the next section we review the Hamilton-Jacobi method for determining the counterterm action, based on this observation.

%%%%%%%%%%%%%%%%%%%%%%%%%%%%%%%%%%%%%%%%%%%%%%%%%%%%%%%%%%%%%
\subsection{The Hamilton-Jacobi Method}
\label{sec:HJMethod}
%%%%%%%%%%%%%%%%%%%%%%%%%%%%%%%%%%%%%%%%%%%%%%%%%%%%%%%%%%%%%

The Hamilton-Jacobi method obtains the counterterm action by solving the constraints associated with diffeomorphism invariance of the gravitational theory \cite{deBoer:1999xf,deBoer:2000cz,Kalkkinen:2001vg}. An especially thorough review of this technique can be found in \cite{Martelli:2002sp}. In this section we will review the standard application of this method, which leads to the same minimal subtraction counterterm action that was found in section \ref{sec:BoundaryTermsasCounterterms}. Then, in section \ref{sec:CountertermHamiltonian}, we will modify this approach to produce a finite, diffeomorphism invariant action.

Before explaining the derivation of the counterterm action it is useful to review the constraints associated with diffeomorphism invariance. In section \ref{sec:OnShell} the on-shell action was regulated by introducing an infra-red cut-off on the coordinate $r$, singling out this coordinate for special treatment. If $\HH$ generates reparameterizations of $r$, and $\HH_{a}$ are the generators of diffeomorphisms of the boundary coordinates $x^{a}$, then diffeomorphism invariance requires
\begin{equation}\label{DiffeoConstraints}
   \HH = 0 \bigsp \bigsp \HH_{a} = 0 ~.
\end{equation}
The coordinate $r$ is playing the same role as time in the standard Hamiltonian formalism, so we will refer to $\HH$ as the Hamiltonian.

In the Hamiltonian formulation of gravity the basic variables are the induced metric $\gamma_{ab}$ on the hypersurface $\Sigma_{r}$, and its conjugate momentum $\pi^{ab}$. In terms of these variables $\HH$ and $\HH_{a}$ are given by
\begin{eqnarray} \label{Hamiltonian}
   \HH  & = & 2 \kappa^{2} \, \sqrt{\gamma}\left(\frac{1}{d-1}\,\pi^{a}_{\,a}\pi^{b}_{\,b} - \pi^{ab}\pi_{ab}\right) -
   \frac{\sqrt{\gamma}}{2\,\kappa^{2}} \,\left(\RR-2\Lambda\right) \\
\label{MomentumGenerator}
    \HH_{a} & = & 2 \sqrt{\gamma}\,\DD^{b}\pi_{ab}  ~.                
\end{eqnarray}
The momentum is given by the derivative of the bulk Lagrangian density\footnote{Given a foliation of the space $\MM$, such as the one presented in section \ref{sec:Foliations}, the contracted Gauss-Codazzi equations can be used to rewrite the action  \eqref{GravityAction} as a bulk integral with no boundary term. This form of the action, familiar from the
ADM construction, is what we refer to here.} with respect to the normal derivative of $\gamma_{ab}$
\begin{equation}\label{momentum}
  \pi^{ab} = \frac{\partial \LL_{\MM}}{\partial (n^{\mu}\partial_{\mu}\gamma_{ab})} 
      = \frac{1}{2\kappa^{2}}\,\left( K^{ab} - \gamma^{ab}\,K\right)  ~.
\end{equation}
Hamilton-Jacobi theory tells us that this momentum, evaluated at the boundary, can also be written as the functional derivative of the on-shell action with respect to the metric on the boundary
\begin{equation}\label{HJmomentum}
    \left.\pi^{ab}\right|_{\Sigma_c} = \left.\frac{1}{\sqrt{\gamma}} 
                       \,\frac{\delta I}{\delta \gamma_{ab}}\right|_{\Sigma_c}  ~.
\end{equation}
This last result is easily verified using the variation of the action under a small change in the metric, given in \eqref{ActionVariation}. Using this expression in \eqref{Hamiltonian} and \eqref{MomentumGenerator} yields a set of functional differential equations for the on-shell action
\begin{equation}
\label{HJequation} 
   \frac{2 \kappa^{2}}{\sqrt{\gamma}} \, \left( \frac{1}{d-1}\,\gamma_{ab}\gamma_{cd}-\gamma_{ac}\gamma_{bd}\right)\,
    \frac{\delta I}{\delta \gamma_{ab}} \, \frac{\delta I}{\delta \gamma_{cd}} - \frac{\sqrt{\gamma}}{2 \kappa^{2}}
    \,\left(\RR-2\Lambda\right)   =   0
\end{equation}
\begin{equation}
\label{MomentumEquation}
 D_{b} \left(  \frac{\delta I}{\delta \gamma_{ab}}\right)  =  0  ~.   
\end{equation}
The first equation is known as the Hamilton-Jacobi equation. It will be our focus in the remainder of this section.

The Hamilton-Jacobi equation is a non-linear functional differential equation and cannot be solved in general. Luckily, the results of the previous section suggest an ansatz for the solution we are looking for. The action is decomposed into renormalized and counterterm contributions
\begin{equation}
\label{SplitAction}
    I =  \Gamma + I_{CT} ~.
\end{equation}
The counterterm action $I_{CT}$ will consist of a finite number of local functionals, intrinsic to the boundary, which should contain the infra-red divergent terms of the on-shell action. Splitting the action into two distinct terms allows us to rewrite the Hamilton-Jacobi equation in terms of the functional derivatives of $\Gamma$ and $I_{CT}$
\begin{equation}\label{OtherMomenta}
   Z^{ab} = \frac{1}{\sqrt{\gamma}}\,\frac{\delta \Gamma}{\delta \gamma_{ab}} \bigsp \bigsp %\bigsp
   P^{ab} = \frac{1}{\sqrt{\gamma}}\,\frac{\delta I_{CT}}{\delta \gamma_{ab}}  ~.
\end{equation}
According to \eqref{SplitAction} these are related to $\pi^{ab}$ by
\begin{equation}\label{MomentumDecomposition}
   \pi^{ab} = Z^{ab} + P^{ab}  ~.
\end{equation}
The Hamilton-Jacobi equation can now be written in the following form
\begin{equation}\label{NewHJ}
  \HH_{CT} + F_{1}[\Gamma] + F_{2}[\Gamma,I_{CT}] = 0  ~.
\end{equation}
The first term is the expression \eqref{HJequation} with $I$ replaced by $I_{CT}$. It will be referred to as the `counterterm Hamiltonian', and is given by
 \begin{equation}\label{CTHJ}
 \HH_{CT}   =  2 \kappa^{2} \, \sqrt{\gamma}\left(\frac{1}{d-1}\, P^{a}_{\,a} P^{b}_{\,b} - P^{ab} P_{ab}\right) -
   \frac{\sqrt{\gamma}}{2\,\kappa^{2}} \,\left(\RR-2\Lambda\right) ~.
\end{equation}
The remaining terms in the Hamilton-Jacobi equation are functionals of $\Gamma$ and $I_{CT}$
\begin{eqnarray} \label{F1}
F_{1}\left[\Gamma\right]  & = & 2\kappa^{2}\,\sqrt{\gamma}\,\left(\frac{1}{d-1}\,Z^{a}_{\,a}\,Z^{b}_{\,b}-
Z^{ab}Z_{ab}\right) \\ \label{F2}
F_{2}\left[\Gamma,I_{CT} \right] & = & 4\kappa^{2}\,\sqrt{\gamma}\,\left(\frac{1}{d-1}\,P^{a}_{\,a}\,Z^{b}_{\,b}- P^{ab}Z_{ab}\right)  ~.
\end{eqnarray}
Obtaining the form \eqref{NewHJ} for the Hamilton-Jacobi equation only made use of the decomposition
\eqref{SplitAction} for the action. Next, we assume the same inverse metric expansion for $I_{CT}$ that was used in section \ref{sec:BoundaryTermsasCounterterms}~\footnote{The inverse metric expansion can also be thought of as a derivative expansion, since the factors of the inverse metric are contracting indices on the boundary metric and its derivatives. Thus, a term with $n$ factors of the inverse metric typically involves $2n$ derivatives of the boundary metric. This approach to solving the Hamilton-Jacobi equation in GR was studied extensively by Salopek and collaborators \cite{Salopek:1990jq,Salopek:1992qy,Salopek:1993un,Parry:mw,Salopek:1994sq}.}. Starting with the expression \eqref{1stCTexpansion} for the counterterm action, the functional derivative that appears in $\HH_{CT}$ is
\begin{equation}
  P^{ab} = P^{ab}_{(0)} + P^{ab}_{(2)} + P^{ab}_{(4)} + \ldots
\end{equation}
The term in this expansion with subscript $(2n)$ comes from the functional derivative of the term in
\eqref{1stCTexpansion} with $n$ factors of the inverse metric. As such, it contains $n+1$ factors of the inverse metric. Using the basis \eqref{LCTbasis} for the terms in the counterterm action, the first few terms in the inverse metric expansion of $P^{ab}$ are
\begin{eqnarray}\label{CTmomentum}
P^{ab}_{(0)} & = & \frac{1}{2}\,\gamma^{ab}\,c_{0} \\ \nonumber
P^{ab}_{(2)} & = & -c_{1}\,\left(\RR^{ab} - \frac{1}{2}\,\gamma^{ab}\RR\right) \\ \nonumber
P^{ab}_{(4)} & = & \frac{1}{2} \, \gamma^{ab}\, \left( c_{2}\,\RR^{2} + c_{3}\, \RR^{ab}\RR_{ab} + c_{4}\,
           \RR^{abcd}\RR_{abcd}\right)  -2 \,c_{2}\RR \RR^{ab} \\ \nonumber 
      & & + \left(2\, c_{2} + c_3 + 2\,c_4 \right)\,\DD^{a}\DD^{b}\RR 
          - \left( 2\,c_2 + \frac{1}{2}\,c_3 \right) \, \gamma^{ab}\,\DD^2 \RR + 4\,c_4\,\RR^{a}_{\,\,c}\,\RR^{bc} 
          \\ \nonumber
      & & - \left(2\,c_2+4\,c_4\right)\,\RR^{acbd}\,\RR_{cd} - \left(c_3 + 4\,c_4\right) \, \DD^2 \RR^{ab}
          - 2 \,c_4 \, \RR^{acde}\,\RR^{b}_{\,\,cde} 
\end{eqnarray} 
Using these expressions in \eqref{CTHJ} results in an inverse metric expansion for $\HH_{CT}$
\begin{equation} \label{HCTinvmetricexpansion}
   \HH_{CT} = \HH_{CT}^{(0)} + \HH_{CT}^{(2)}+ \HH_{CT}^{(4)} + \ldots
\end{equation}
where a superscript $(2n)$ indicates $n$ factors of the inverse metric. The terms appearing at the first three orders in \eqref{HCTinvmetricexpansion} are
\begin{eqnarray} \label{HCT0}
  \HH_{CT}^{(0)} & = & \frac{d}{2(d-1)}\,\kappa^2\,\sqrt{\gamma}\,\left( c_{0}^{\,2} - \frac{(d-1)^2}{\kappa^4 \, 
     \ell^2}\right)  \\ \label{HCT2}
  \HH_{CT}^{(2)} & = & \kappa^2 \, \sqrt{\gamma}\,\left( \left(\frac{d-2}{d-1}\right)\,c_0 \, c_1 - 
     \frac{1}{2\,\kappa^4}\right) \, \RR  \raisebox{20pt}{\,}   \\ \label{HCT4}
  \HH_{CT}^{(4)} & = & 2 \kappa^2 \, \sqrt{\gamma}\,\left( \frac{d}{2(d-1)}\,c_{1}^{\,2} + 
      \frac{(d-4)}{4(d-1)}\,c_0 \, c_2 \right) \, \RR^2  \raisebox{20pt}{\,} \\ \nonumber  
      &  & +  \raisebox{20pt}{\,} 2 \, \kappa^2 \, \sqrt{\gamma} \, \left( \frac{(d-4)}{4(d-1)}\,c_0 \, c_3
           - 2 \,c_1^{\,2} \right)\,\RR^{ab}\RR_{ab} \\ \nonumber   
     &   & - \raisebox{20pt}{\,} 2\kappa^2 \,\sqrt{\gamma} \,  c_{0}\,\left( c_{2} + \frac{d}{4(d-1)}\,c_{3}
           + \frac{1}{d-1} \, c_4 \right) \,\DD^{2}\RR \\ \nonumber
   &   & + \raisebox{20pt}{\,} \, \kappa^2\,\sqrt{\gamma}\,\frac{(d-4)}{2(d-1)}\,c_0 \, c_4 \, 
         \RR^{abcd}\,\RR_{abcd} ~.
\end{eqnarray}
The ansatz \eqref{1stCTexpansion} for the counterterm action has led to an expression for the Hamilton-Jacobi
equation
\begin{equation}\label{2ndHJ}
    \HH_{CT}^{(0)} + \HH_{CT}^{(2)} + \ldots + F_{1}\left[\Gamma\right] + F_{2}\left[\Gamma, I_{CT}\right] = 0
\end{equation}
that should be solved order-by-order in the inverse metric expansion. 

When the dimension of the boundary is even, $d=2k$, the first $k-1$ terms in \eqref{2ndHJ} scale as different (positive) powers of the cut-off and must be set to zero. This gives a set of `descent equations' \cite{Martelli:2002sp}
\begin{equation}\label{DescentEq}
    \HH_{CT}^{(2n)} = 0 \bigsp \bigsp 2n < d  ~.
\end{equation}
These equations determine the coefficients for terms in the counterterm action with up to $k-1$ factors of the inverse metric. The term in the expansion involving $k$ factors of the inverse metric, $\HH_{CT}^{(2k)}$, is finite. At this order the descent equations break down, because the functionals $F_1$ and $F_2$ in the Hamilton-Jacobi equation contain terms which are finite. The remaining part of the Hamilton-Jacobi equation is
\begin{equation}\label{FiniteHJTerm}
    \HH_{CT}^{(2k)} + F_{1}\left[\Gamma\right] + F_{2}\left[\Gamma, I_{CT}\right] = 0  ~.
\end{equation}
Unless some additional ansatz is specified, only the divergent terms in the counterterm action are determined by this procedure. Thus, solving the descent equations \eqref{DescentEq} yields the same minimal subtraction counterterm action \eqref{MSCTaction} that was found in section \ref{sec:BoundaryTermsasCounterterms}. Once again there is a natural ambiguity corresponding to a finite term in $I_{CT}$, which is left unresolved due to the break-down in the descent equations. Note that when the boundary dimension is odd there is no finite term in \eqref{HCTinvmetricexpansion}, and the counterterm action is completely determined by this method.

%%%%%%%%%%%%%%%%%%%%%%%%%%%%%%%%%%%%%%%%%%%%%%%%%%%%%%%%%%%%%
\subsection{Minimal Subtraction and Diffeomorphisms}
\label{sec:CountertermsAndDiffeos}
%%%%%%%%%%%%%%%%%%%%%%%%%%%%%%%%%%%%%%%%%%%%%%%%%%%%%%%%%%%%%

Working with the on-shell action requires the introduction of an infra-red cut-off on the coordinate $r$. The cut-off renders the action finite but it does so at the expense of diffeomorphism invariance; reparameterizations of the coordinate normal to the boundary are manifestly broken by the choice of regulator. Of course, once the divergent terms in the action have been cancelled the cut-off can be removed, leaving a finite action $\Gamma_{MS}$. But the fact that diffeomorphism invariance was not preserved at all stages in the renormalization procedure makes it unclear whether or not $\Gamma_{MS}$ really exhibits this symmetry. An explicit calculation shows that it does not.

As stated in section \ref{sec:Divergences}, the boundary of an asymptotically $AdS_{2k+1}$ spacetime possesses a conformal class of metrics. The choice of a non-covariant regulator identifies a particular representative from the conformal class \cite{Graham:1999pm} corresponding to the limit \eqref{InducedRepRelation} of the induced metric on the boundary of the regulated space. Depending on the geometry of the boundary\footnote{The definition of an asymptotically $AdS$ spacetime fixes the topology of the boundary to be $\BR \times S^{d-1}$. For the asymptotically locally $AdS$ spacetimes discussed in \cite{Papadimitriou:2005ii} additional factors, such as the topology of the boundary and choice of conformal structure, also play a role.}, the renormalized action $\Gamma_{MS}$ may retain information about this choice of conformal representative, even after the cut-off has been removed. From the point of view of the AdS/CFT correspondence this represents a conformal anomaly in the dual field theory. The relationship between conformal anomalies and diffeomorphisms can be seen by considering a transformation of the bulk coordinates: $x^{\mu} \to x^{\mu} - \eps^{\mu}(x)$. The renormalized action transforms as \cite{Papadimitriou:2005ii}
\begin{equation}\label{AnomalousTransform}
  \delta_{\eps}\Gamma_{MS} \sim \int_{\dM}\bns d^{d}x \, \sqrt{\hat{h}}\,\AA[\hat{h}]\, n_{\mu}\,\eps^{\mu}(x) ~.
\end{equation}
where $\hat{h}_{ab}$ is the conformal representative and $\AA[\hat{h}]$ is the conformal anomaly density of the dual field theory. Note that this term is finite by power-counting\footnote{The power counting rules are only sensitive to power-law dependence on the cut-off. But at leading order, the normal component of the diffeomorphism may in fact scale as the log of the cut-off. We ignore this additional complication for the moment and focus on the fact that \eqref{AnomalousTransform} does not vanish for generic asymptotically $AdS$ spacetimes. In later sections of this paper the potential log divergence of the normal component of the diffeomorphism will not pose a problem.}. The normal component of the diffeomorphism $\eps^{\mu}$ may have an arbitrary dependence on the boundary coordinates, so the action $\Gamma_{MS}$ is only invariant under bulk diffeomorphisms if the anomaly density vanishes. This is true for global $AdS$ or the $AdS$-Schwarzschild solution, but not for the Kerr-$AdS_{d+1}$ spacetime, whose renormalized action was calculated using minimal subtraction in \cite{Awad:1999xx,Awad:2000ac,Awad:2000aj}. Therefore, diffeomorphism invariance is not a general property of the action obtained using minimal subtraction.

Upon closer inspection, it seems that the failure of diffeomorphism invariance is related to the breakdown in the descent equations \eqref{DescentEq}. By direct calculation \cite{Martelli:2002sp},
the term $\HH_{CT}^{(2k)}$ is proportional to the anomaly density of the dual field theory
\begin{equation}
   \HH_{CT}^{(2k)} \sim \sqrt{\hat{h}}\,\AA[\hat{h}] ~.
\end{equation}
The transformation \eqref{AnomalousTransform} of the renormalized action $\Gamma_{MS}$ can now be expressed as
\begin{equation}\label{AnomalousTransformHamiltonian}
   \delta_{\eps}\,\Gamma_{MS} \sim \int_{\dM}\bns d^{d}x \,\HH_{CT}^{(2k)} \,
       n_{\mu}\,\eps^{\mu}  ~.
\end{equation}
It seems reasonable, if our goal is to obtain a renormalized action that is invariant under bulk diffeomorphisms, to try to modify the counterterm action so that 
\begin{equation}
      \HH_{CT}^{(2k)} = 0
\end{equation}
when the boundary dimension is $d=2k$. In the next section we will show that this can be accomplished for
asymptotically $AdS$ spacetimes.

%%%%%%%%%%%%%%%%%%%%%%%%%%%%%%%%%%%%%%%%%%%%%%%%%%%%%%%%%%%%%
%%%%%%%%%%%%%%%%%%%%%%%%%%%%%%%%%%%%%%%%%%%%%%%%%%%%%%%%%%%%%
\section{Modifying The Counterterm Action}
\label{sec:CountertermHamiltonian}
%%%%%%%%%%%%%%%%%%%%%%%%%%%%%%%%%%%%%%%%%%%%%%%%%%%%%%%%%%%%%
%%%%%%%%%%%%%%%%%%%%%%%%%%%%%%%%%%%%%%%%%%%%%%%%%%%%%%%%%%%%%

In the previous section we saw that the renormalized action obtained via minimal subtraction is not invariant under certain diffeomorphisms when the dimension of the boundary is even. Explicitly working out the transformation properties of $\Gamma_{MS}$ under a diffeomorphism shows that this is related to the breakdown in the descent equations \eqref{DescentEq}. In this section we show that adding the condition $\HH_{CT}^{(2k)} = 0$ to our ansatz for solving the Hamilton-Jacobi equation leads to a new renormalized action, $\Gamma_{HJ}$, that is invariant under diffeomorphisms. Imposing this condition requires the introduction of a new counterterm which is finite by power-counting. The role of this term is apparent in dimensional regularization, as is the proof of its existence for asymptotically $AdS_{2k+1}$ spacetimes.

%%%%%%%%%%%%%%%%%%%%%%%%%%%%%%%%%%%%%%%%%%%%%%%%%%%%%%%%%%%%%
\subsection{A New Counterterm}
\label{sec:ANewCounterterm}
%%%%%%%%%%%%%%%%%%%%%%%%%%%%%%%%%%%%%%%%%%%%%%%%%%%%%%%%%%%%%

Solving the descent equations \eqref{DescentEq} for arbitrary $d$ yields the following inverse metric
expansion for the counterterm action
\begin{eqnarray}
\label{CTaction2}
  I_{CT} & = & \frac{1}{\kappa^{2}} \, \int_{\Sigma_c} \nts \nts d^{d}x \, \sqrt{\gamma} \left(\frac{d-1}{\ell} + \frac{\ell}{2(d-2)}
                   \,\RR \right. \\ \nonumber
             &     &        \hspace{2cm}\left. + \frac{\ell^3}{2(d-2)^2 (d-4)}\left( \RR^{ab} \RR_{ab} -
                   \frac{d}{4(d-1)}\,\RR^2\right)  + \ldots \right)  ~.
\end{eqnarray}
If we accept that the descent equations break down with the finite term \eqref{FiniteHJTerm} then the
Hamilton-Jacobi approach gives the same counterterm action as minimal subtraction. On the other hand, if we let $d$ take the non-integral value $2k+ 2\varepsilon $, with $\varepsilon$ small and positive, then there is no breakdown in the descent equations. Rather, the term $\HH_{CT}^{(2k)}$ appearing in the inverse metric expansion of $\HH_{CT}$ scales as a positive power of the cut-off and must be cancelled by additional terms in the counterterm action. To do this we include the terms in \eqref{CTaction2} at the next order beyond what is required by minimal subtraction. The counterterm action associated with our new solution of the Hamilton-Jacobi equation is then given by
\begin{equation}\label{HJaction}
   I_{HJ}  =  \frac{1}{\kappa^{2}} \, \int_{\Sigma_c} \nts \nts d^{2k+2\varepsilon}x \, \sqrt{\gamma} \,\sum_{n=0}^{k} \LL_{CT}^{(2n)}
\end{equation}
and the new renormalized action is
\begin{equation}\label{NewRenormalizedAction}
  \Gamma_{HJ} = I - I_{HJ} ~.
\end{equation}
The counterterm action is related to the minimal subtraction counterterm action by
\begin{equation}
    I_{HJ} = I_{MS} + I_{CT}^{(2k)}
\end{equation}
where the additional term corresponds to
\begin{equation}\label{NewCT1}
   I_{CT}^{(2k)} = \frac{1}{\kappa^2}\,\int_{\Sigma_c} \nts \nts d^{2k+2\varepsilon}x \, \sqrt{\gamma}\,\LL_{CT}^{(2k)}~.
\end{equation}
It is important to remember that this counterterm depends on the dimension $d=2k+2\varepsilon$. This means that, in addition to evaluating the integrand away from $d=2k$, the integral must be properly generalized to a $2k+2\varepsilon$ dimensional integral. The new contribution to the counterterm action is then obtained by taking the $d \to 2k^+$, or $\varepsilon \to 0^+$, limit of \eqref{NewCT1}.

Inspecting the counterterm action \eqref{CTaction2} reveals an immediate problem. The term $I_{CT}^{(2k)}$
appears to contain a pole in the $d \to 2k^+$ limit. For instance, if $d=4 + 2\varepsilon$ then
\begin{equation}\label{FourdFiniteTerm}
    I_{CT}^{(4)} = \lim_{d \to 4^{+}} \frac{1}{\kappa^{2}}\,\int_{\Sigma_c} \nts \nts d^{d}x \, \sqrt{\gamma}\,
        \frac{\ell^{3}}{2(d-2)^{2}(d-4)}\,\left(\RR^{ab}\RR_{ab}-\frac{d}{4(d-1)}\,\RR^{2}\right) ~.
\end{equation}
The pole appears because the descent equation $\HH_{CT}^{(4)} = 0$ involves terms with factors of $(d-4)$
in their coefficients, as can be seen in \eqref{HCT4}. However, in a moment we will show that the $d \to 4^+$ limit of $I_{CT}^{(4)}$ is perfectly well defined when the bulk spacetime is asymptotically $AdS_5$. More generally, the $d \to 2k^+$ limit of the counterterm $I_{CT}^{(2k)}$ exists  for all asymptotically $AdS_{2k+1}$ spacetimes. The proof of this statement depends on the properties that were emphasized at the end of section \ref{sec:Conventions}. 

Before proving the existence of the $d \to 2k^+$ limit of $I_{CT}^{(2k)}$, it is useful to see a simple example worked out in detail. Let $\MM$ be an asymptotically $AdS_{5}$ spacetime whose boundary $\dM$ has geometry $\BR \times S^{3}$. In this case, the generalization to a $d= 4 + 2\varepsilon$ dimensional boundary is straightforward
\begin{equation}
    \BR \times S^{3} \rightarrow \BR \times S^{(d-1)} ~.
\end{equation}
The intrinsic curvature tensors that appear in \eqref{FourdFiniteTerm} are easily calculated for this geometry. They are given by
\begin{equation}
   \RR_{ij} = \frac{(d-2)}{r_{c}^{2}}\,\gamma_{ij} \bigsp \bigsp \RR = \frac{(d-1)(d-2)}{r_{c}^{2}}
\end{equation}
where indices $i,j$ represent coordinates on the sphere $S^{d-1}$. Using these expressions the curvature terms in \eqref{FourdFiniteTerm} give
\begin{equation}\label{QuadCurvTerms}
 \RR^{ab}\RR_{ab} - \frac{d}{4\,(d-1)}\,\RR^2 = -\frac{1}{4\,r_{c}^4}\,(d-1)(d-2)^2 (d-4) ~.
\end{equation}
The apparent pole in \eqref{FourdFiniteTerm} is cancelled by an explicit factor of $(d-4)$ coming from the curvature terms. The counterterm can now be directly evaluated. In the $d \to 4^+$ limit the factor of $r_{c}^{-4}$ from  the curvature terms cancels a factor of $r_{c}^d$ coming from the volume element, so that the final result depends only on the conformal representative \eqref{InducedRepRelation}. Before performing the integral we replace the non-compact time coordinate (the factor $\BR$ in the geometry) with an $S^1$ corresponding to periodic imaginary time on the Euclidean section of $\MM$.
The result is
\begin{equation}
   I_{CT}^{(4)} = -\frac{3 \,\ell^{2}}{8\,\kappa^{2}} \,\beta\,\omega_{3}
\end{equation} 
where $\beta$ is the periodicity of the $S^{1}$ and $\omega_{3}$ is the volume of the unit $S^{3}$. Note that this term is equal to the finite difference between the background subtraction and boundary counterterm actions for both the global $AdS_5$ and the Schwarzschild-$AdS_5$ spacetimes.

The previous example is simple because of the high degree of symmetry in the boundary geometry. For less symmetric geometries the computation of the counterterm $I_{CT}^{(2k)}$ is more difficult. But the existence of this term, which we show in the next section, can be shown quite generally.

%%%%%%%%%%%%%%%%%%%%%%%%%%%%%%%%%%%%%%%%%%%%%%%%%%%%%%%%%%%%%
\subsection{Existence of the Counterterm \texorpdfstring{$I_{CT}^{(2k)}$}{} in \texorpdfstring{$d=2k$}{d=2k}}
\label{sec:FiniteCounterterm}
%%%%%%%%%%%%%%%%%%%%%%%%%%%%%%%%%%%%%%%%%%%%%%%%%%%%%%%%%%%%%

As stated in the previous section, we can show that $I_{CT}^{(2k)}$ has a well defined $d \to 2k^+$ limit for all asymptotically $AdS_{2k+1}$ spacetimes. The argument depends on the two properties of asymptotically $AdS$ spacetimes emphasized at the end of section \ref{sec:Conventions}. The first step is to rewrite the counterterm action in a different basis. Instead of using the generic boundary curvature polynomials in \eqref{LCTbasis} we will use curvature polynomials $\EE_{2n}(d)$ that coincide with the Euler density in $d=2n$ dimensions, and the set of polynomials $\CC_{2n}^{(i)}(d)$ that are exactly Weyl invariant in $d=2n$. The first few terms that we will need are
\begin{eqnarray} \label{NewBasis}
   \EE_{2}(d) & = & \frac{1}{4}\,\RR \\ \nonumber 
   \EE_{4}(d) & = & \frac{1}{64}\,\left(\RR^{abcd}\RR_{abcd}-4 \RR^{ab}\RR_{ab}+\RR^{2}\right)   \\ \nonumber
   \CC_{4}(d) & = & -\frac{1}{64}\,\left(
\RR^{abcd}\RR_{abcd}-\frac{4}{d-2}\,\RR^{ab}\RR_{ab}+\frac{2}{(d-1)(d-2)}\,\RR^{2}\right)
\end{eqnarray} 
Higher order terms in this basis can be found, for instance, in \cite{Bonora:1985cq,Henningson:1998gx}. The last term in \eqref{NewBasis} is the unique Weyl invariant polynomial in four dimensions. It is proportional to the square of the Weyl tensor, which by definition may be written in the form above. 

In the new basis the counterterm action can be written as
\begin{eqnarray}
I_{CT} & = & \frac{1}{\kappa^{2}}\,\int_{\Sigma_c} \nts \nts d^{d}x \, \sqrt{\gamma}\,\frac{(d-1)}{\ell}\,\left[ 1 +
\frac{2\,\ell^{2}}{(d-1)(d-2)}\,\EE_{2}(d) \right. \\ \nonumber
           &    & \left.\hspace{2cm} -\frac{8\,\ell^{4}}{(d-1)(d-2)(d-3)(d-4)} \, \left(\raisebox{12pt}{\,}
          \EE_{4}(d) + \CC_{4}(d) \right) 
 + \ldots \right] ~.
\end{eqnarray}
Let us now analyze the counterterm that was encountered in the previous section, for an asymptotically $AdS_5$ spacetime. It takes the form
\begin{equation} \label{TermIn4Dim}
 I_{CT}^{(4)} = -\frac{1}{\kappa^{2}} \, \lim_{d \to 4^+} \int_{\Sigma_c} \nts \nts d^{d}x \, 
      \sqrt{\gamma}\,\frac{8\,\ell^{3}}{(d-2)(d-3)(d-4)} \, \left( \raisebox{12pt}{\,}
      \EE_{4}(d) + \CC_{4}(d) \right) ~.
\end{equation}
As in the previous example, the volume element scales with the cut-off as $r_{c}^{d}$, while the terms that are quadratic in the boundary curvatures scale as $r_{c}^{-4}$. Factoring out this $r_{c}$ dependence, the only terms that do not vanish when the cut-off is removed depend on the conformal representative $\hat{h}_{ab}$
\begin{equation}\label{TermIn4DimNoCutOff}
 I_{CT}^{(4)} = -\frac{1}{\kappa^{2}} \, \lim_{d \to 4^+} \int_{\dM} \bns d^{d}x \, \sqrt{\hat{h}} 
     \,\left(\frac{r_{c}}{\ell}\right)^{2\,\varepsilon}\, \frac{8\,\ell^{3}}{(d-2)(d-3)(d-4)} \, 
     \left(\,\hat{\EE}_{4}(d) +\hat{\CC}_{4}(d) \right) ~.
\end{equation}
The term $\hat{\CC}_{4}(d)$ automatically vanishes, because the Weyl tensor for any representative of the conformal class of metrics on the boundary of an asymptotically $AdS$ spacetime vanishes. The remaining term in the integrand is a quadratic curvature polynomial whose integral in $d=4$ is a topological invariant. But in $d=4+2\varepsilon$ dimensions it contains geometric information as well. Expanding for $d$ near four dimensions, we can write $\hat{\EE}_{4}(d)$ as
\begin{equation}\label{ExpansionOfE}
   \hat{\EE}_{4}(d) = \hat{\EE}_{4}(4) + (d-4)\,\hat{\EE}_{4}^{(1)}(4) + \frac{1}{2}\,(d-4)^2 \, 
    \hat{\EE}_{4}^{(2)}(4) + \ldots
\end{equation}
The example in the previous section showed this expansion explicitly, in equation \eqref{QuadCurvTerms}. The first term is the four dimensional Euler density. As stated in section \ref{sec:Conventions}, the boundary of an asymptotically $AdS$ spacetime has topology $\BR \times S^{d-1}$, which has an Euler number of zero. Therefore, the integral of the first term in the expansion \eqref{ExpansionOfE} vanishes. The next term carries an explicit factor of $(d-4)$ that cancels the $(d-4)^{-1}$ pole in \eqref{TermIn4DimNoCutOff}. Unlike the Euler density, whose integral is topological, this term encodes geometric information and depends on the choice of conformal representative $\hat{h}_{ab}$. Any contributions to \eqref{TermIn4DimNoCutOff} from higher order terms in \eqref{ExpansionOfE} vanish in the $d \to 4^+$ limit. Thus, the term $I_{CT}^{(4)}$ appearing in the counterterm action for an asymptotically $AdS_{5}$ spacetime is given by
\begin{equation}
  I_{CT}^{(4)} = \lim_{d \to 4^+} \frac{1}{\kappa^{2}}\,\frac{8\,\ell^{3}}{(d-2)(d-3)}\,
     \int_{\dM}\bns d^{d}x \, \sqrt{\hat{h}}\,\hat{\EE}_{4}^{(1)}(4) ~.
\end{equation}
This term is finite and well defined in the $ d \to 4^+$ limit.

These arguments are not particular to a four dimensional boundary. For an asymptotically $AdS_{2k+1}$ spacetime the term $I_{CT}^{(2k)}$ will have a form analogous to \eqref{TermIn4DimNoCutOff}
\begin{equation}
 I_{CT}^{(2k)} \sim \lim_{d \to 2k^+} \,\int_{\Sigma_c} \nts \nts d^{d}x \sqrt{\hat{h}}\,\frac{1}{(d-2k)}\,
  \left(\frac{r_{c}}{\ell}\right)^{2\,\varepsilon}\,\left( \, \hat{\EE}_{2k}(d) + \sum_i \alpha_i \, \hat{\CC}_{2k}^{(i)}(d)\right) ~.
\end{equation}
The terms $\hat{\CC}_{2k}^{(i)}$ appearing in the integrand are given by contractions of the Weyl tensor and its derivatives, and therefore vanish. The remaining term will correspond to the curvature polynomial $\hat{\EE}_{2k}(d)$ that gives the Euler density when evaluated at $d=2k$ dimensions. It can be expanded around $d=2k$ as in \eqref{ExpansionOfE}, so that the non-vanishing contribution to $I_{CT}^{(2k)}$ is given by
\begin{equation}
\label{GeneralForm}
  I_{CT}^{(2k)} \sim \lim_{d \to 2k^+} \, \int_{\dM} \bns d^{d}x\,\sqrt{\hat{h}} \, \hat{\EE}_{2k}^{(1)}(2k) ~.
\end{equation}

%%%%%%%%%%%%%%%%%%%%%%%%%%%%%%%%%%%%%%%%%%%%%%%%%%%%%%%%%%%%%
\subsection{Diffeomorphism Invariance of \texorpdfstring{$\Gamma_{HJ}$}{GammaHJ}}
\label{sec:GammaAndDiffeos}
%%%%%%%%%%%%%%%%%%%%%%%%%%%%%%%%%%%%%%%%%%%%%%%%%%%%%%%%%%%%%

In the previous section we showed that the counterterm $I_{CT}^{(2k)}$ is well defined for asymptotically $AdS$ spacetimes of dimension $d+1=2k+1$ . Including this term in the counterterm action guarantees that the term $\HH_{CT}^{(2k)}$ in the counterterm Hamiltonian vanishes. We will now show that, as a result of this modification, the renormalized action $\Gamma_{HJ}$ is invariant under bulk diffeomorphisms. 

Consider an infinitesimal reparameterization of the bulk coordinates
\begin{equation}\label{diffeo}
   x^{\mu} \rightarrow x^{\mu} - \eps^{\mu}(x) ~.
\end{equation}
We can study the behavior of $\Gamma_{HJ}$ under this diffeomorphism using the well-defined transformation properties of the on-shell action $I$ and the counterterm action $I_{HJ}$
\begin{equation}
   \delta_{\eps} \Gamma_{HJ} = \delta_{\eps} I - \delta_{\eps} I_{HJ} ~.
\end{equation}
The transformation of the action $I$ can be obtained from \eqref{ActionVariation} after noting that \eqref{diffeo} changes the bulk metric according to
\begin{equation}
   \delta_{\eps} g_{\mu\nu} = \nabla_{\mu}\,\eps_{\nu} + \nabla_{\nu}\,\eps_{\mu} ~.
\end{equation}
After an integration by parts, the transformation of the regulated action is given by
\begin{equation}
\delta_{\eps} I = - \frac{1}{\kappa^2}\,\int_{\MM_{c}} \nts \nts d^{d+1}x \sqrt{g} \, \eps_{\nu} \nabla_{\mu} G^{\mu\nu}
 +  \int_{\Sigma_c} \nts \nts d^{d}x \sqrt{\gamma} \, \left( \frac{1}{\kappa^2} \, G^{\mu\nu} n_{\mu} \eps_{\nu} 
   + \pi^{ab} \, \delta_{\eps} \gamma_{ab} \right)
\end{equation}
where $G_{\mu\nu}$ is the $d+1$-dimensional Einstein tensor, including the contribution from the cosmological constant.
The bulk term vanishes automatically by the twice-contracted Bianchi identity. The first boundary term can be simplified by splitting the diffeomorphism parameter $\eps^{\mu}$ into its normal and boundary components. The resulting projections of the Einstein tensor are proportional to $\HH$ and $\HH_a$. The second boundary term involves the variation of the boundary metric under \eqref{diffeo}. The induced metric on the boundary transforms as a tensor under diffeomorphisms of the boundary coordinates, but it is simply a scalar from the point of view of the coordinate normal to the boundary. Therefore its variation is
\begin{equation}
  \delta_{\eps}\gamma_{ab} = \DD_a \eps_b + \DD_b \eps_a + n^{\mu}\,\partial_{\mu} \gamma_{ab}
\end{equation}
The last term in this expression is proportional to the extrinsic curvature of the boundary, using \eqref{ExtrinsicCurvature2}. Thus, the change in the regulated on-shell action under a bulk diffeomorphism \eqref{diffeo} is
\begin{equation}\label{ITransform}
\delta_{\eps} I = \int_{\Sigma_c} \nts \nts d^{d}x \sqrt{\gamma} \left[ \frac{1}{\kappa^2}
  \,\left( \HH\, n_{\mu} \eps^{\mu}  + \HH_{a} \, \eps^a \right) 
   + 2 \, \pi^{ab} K_{ab} \, n_{\mu} \eps^{\mu} + 2 \, \pi^{ab} \DD_a \eps_b \right] ~.
\end{equation}

The counterterm action is a local functional of the induced metric on the boundary, so its transformation properties under \eqref{diffeo} are very simple
\begin{equation}
   \delta_{\eps} I_{HJ} = \int_{\Sigma_c} \nts \nts d^{d}x \, \frac{\delta I_{HJ}}{\delta \gamma_{ab}} \, \delta_{\eps} 
     \gamma_{ab} ~.
\end{equation}
Using the definition \eqref{OtherMomenta} we can replace the functional derivative of $I_{CT}$ with $\sqrt{\gamma}\,P^{ab}$. This expression, along with \eqref{ITransform}, allows us to write the transformation of the renormalized action $\Gamma_{HJ}$ under a bulk diffeomorphism. After an integration by parts, we arrive at the following result
\begin{equation}\label{GammaTransform}
\delta_{\eps} \Gamma_{HJ} = \int_{\Sigma_c} \nts \nts d^{d}x \sqrt{\gamma} 
    \left[ \frac{1}{\kappa^2}\,\left( \HH\, n_{\mu}\eps^{\mu} + \HH_{a} \, \eps^a \right)  
    - \eps_{a}\,\DD_{b} T^{ab} +  T^{ab} K_{ab} \, n_{\mu} \eps^{\mu} 
   \right]
\end{equation}
where we have defined
\begin{equation}
  T^{ab} = 2 \left( \pi^{ab} - P^{ab}\right) ~.
\end{equation}
Note that the quantity $T^{ab}$ is just the quasi-local stress tensor of Brown and York \cite{Brown:1992br} for the action $\Gamma_{HJ}$. 

At this point the regulator is still in place; $\Sigma_c$ is the hypersurface defined by the infra-red cut-off. For the renormalized action to be invariant under bulk diffeomorphisms we must show that the expression \eqref{GammaTransform} is zero, up to terms which vanish as the cut-off is removed, for arbitrary $\epsilon^{\mu}$. It is straightforward to show that the first three terms in \eqref{GammaTransform} are zero. The first two terms are proportional to $\HH$ and $\HH_a$. They vanish on-shell because they are simply projections of the bulk Einstein tensor. The third term involves the (boundary) covariant divergence of the boundary stress tensor. To see that this must vanish, first note that it is related to the renormalized stress tensor obtained from minimal subtraction \cite{Balasubramanian:1999re} by
\begin{equation}\label{NewStressTensorDivergence}
   \DD_{b} T^{ab} = \DD_{b} T^{ab}_{MS} - \DD_{b} P^{ab}_{(2k)} ~.
\end{equation}
The fact that the first term vanishes was established in \cite{Balasubramanian:1999re}. The second term vanishes as well, by repeated application of the contracted and uncontracted Bianchi identities. For instance, when $2k=2$ this amounts to
\begin{equation}
   \DD_a P^{ab}_{(2)} \sim \DD_a \GG^{ab}
\end{equation}
where $\GG^{ab}$ is the boundary Einstein tensor. The divergence of $\GG^{ab}$ vanishes by application of the twice-contracted Bianchi identity. Verifying that the divergence of $P^{ab}_{(4)}$ vanishes requires both the once- and twice-contracted forms of the identity. Although the form of the second term in \eqref{NewStressTensorDivergence} becomes much more complicated for $2k \geq 6$, the same arguments can be used to show that it vanishes in those cases.

Showing that the last term in \eqref{GammaTransform} vanishes is more involved. Specifically, we want to show that
\begin{equation}\label{normalterm}
   \sqrt{\gamma}\,T^{ab}\,K_{ab} = 0 + \ldots
\end{equation}
where `$\ldots$' denote terms which vanish as the cut-off is removed. First, it is useful to rewrite the left hand side of \eqref{normalterm}. Using \eqref{momentum} and \eqref{MomentumDecomposition} the Hamiltonian \eqref{Hamiltonian} can be rewritten in the following form
\begin{equation}
\HH = -\sqrt{\gamma} \, K_{ab}\,\left( P^{ab} + \frac{1}{2} \, T^{ab} \right) - 
    \frac{\sqrt{\gamma}}{2 \kappa^2} \, \left( \RR - 2 \Lambda \right)  ~.
\end{equation}
The Hamiltonian constraint then implies
\begin{equation}
  \sqrt{\gamma} \, T^{ab}\, K_{ab} = -\sqrt{\gamma}\,2\,K_{ab}P^{ab}-\frac{\sqrt{\gamma}}{\kappa^2}\,\left( 
   \RR-2 \Lambda\right) ~.
\end{equation}
We can now explicitly show that the right hand side of this equation vanishes, using asymptotic expansions for the intrinsic and extrinsic curvatures. This is described in detail for the case of asymptotically $AdS_5$ spacetimes in appendix \ref{app:AsymptoticExpansions}. The result is that \eqref{normalterm} is satisfied, so that
\begin{equation}
  \delta_{\eps} \Gamma_{HJ} = 0 ~.
\end{equation}
We have found that including $I_{CT}^{(2k)}$ in the counterterm action leads to a renormalized action which is invariant under bulk diffeomorphisms.

%%%%%%%%%%%%%%%%%%%%%%%%%%%%%%%%%%%%%%%%%%%%%%%%%%%%%%%%%%%%%
\section{The Kerr-\texorpdfstring{$AdS_{5}$}{AdS5} Black Hole}
\label{sec:KerrAdS5}
%%%%%%%%%%%%%%%%%%%%%%%%%%%%%%%%%%%%%%%%%%%%%%%%%%%%%%%%%%%%%

In this section we review the properties of the Kerr-$AdS_{5}$ black hole and summarize the calculation of the action using background subtraction. We then calculate the renormalized action using the techniques discussed in sections \ref{sec:BoundaryCounterterms} and \ref{sec:CountertermHamiltonian}. We find that our modified version of the boundary counterterm approach gives the same renormalized action and conserved charges as background subtraction.

The metric for Kerr-$AdS_{d+1}$ is discussed in Appendix \ref{app:GeneralKerrAdS}. In 5 dimensions it can be written as
\begin{eqnarray}
\label{KerrAdS5Metric}
 ds^{2} & = & \frac{U}{V-2m}\,dr^{2} - W\,\left(1+\frac{r^{2}}{\ell^{2}}\right) dt^{2}
    +\frac{2m}{U}\left(dt - \frac{a}{\Xi_{a}}\mu_{1}^{\,2}d\phi - \frac{b}{\Xi_{b}}\mu_{2}^{\,2}d\psi\right)^{2} \\ \nonumber
  & & + \,\frac{r^{2}+a^{2}}{\Xi_{a}}\,\mu_{1}^{\,2}\,\left(d\phi + \frac{a}{\ell^{2}}\,dt\right)^{2}
          + \,\frac{r^{2}+b^{2}}{\Xi_{b}}\,\mu_{2}^{\,2}\,\left(d\psi + \frac{b}{\ell^{2}}\,dt\right)^{2} \\ \nonumber
  & & + \,\left(\frac{r^{2}+a^{2}}{\Xi_{a}}+\frac{r^{2}+b^{2}}{\Xi_{b}}\,\frac{\mu_{1}^{\,2}}{\mu_{2}^{\,2}}
        -\frac{1}{\ell^{2}\,W}\,\left(1+\frac{r^{2}}{\ell^{2}}\right)\,
        \frac{(\Xi_{a}-\Xi_{b})^{2}}{\Xi_{a}^{\,2}\,\Xi_{b}^{\,2}} \,\mu_{1}^{\,2}\right)\,d\mu_{1}^{\,2} ~.
\end{eqnarray}
The functions appearing in the metric are
\begin{eqnarray}
   W & = & \frac{\mu_{1}^{\,2}}{\Xi_{a}} + \frac{\mu_{2}^{\,2}}{\Xi_{b}} \\
   U & = & r^{2} +a^{2}\,\mu_{2}^{\,2}+b^{2}\,\mu_{1}^{\,2} \\
   V & = & \frac{1}{r^{2}}\,\left(1+\frac{r^{2}}{\ell^{2}}\right)\,(r^{2}+a^{2})\,(r^{2}+b^{2}) ~.
\end{eqnarray}
A single angle $\theta \in [0,\pi/2]$ parameterizes the two direction cosines as $\mu_{1} = \sin(\theta)$ and
$\mu_{2}=\cos(\theta)$. There are two independent rotation parameters, which we denote by $a$ and $b$, and two
corresponding azimuthal angles, $\phi$ and $\psi$, which run over $[0,2\pi]$. We define an angular velocity in each plane of rotation, relative to a non-rotating frame at infinity, by
\begin{equation} \label{AVKerrAdS5}
  \Omega_{a} = \frac{a}{r_{h}^{\,2}+a^{2}}\,\left(1+\frac{r_{h}^{\,2}}{\ell^{2}}\right) \hspace{1cm}
    \Omega_{b} = \frac{b}{r_{h}^{\,2}+b^{2}}\,\left(1+\frac{r_{h}^{\,2}}{\ell^{2}}\right) ~.
\end{equation}
Using \eqref{HorizonArea} the area of the horizon is
\begin{equation}
   A_{h} = \frac{2 \pi^{2}}{r_{h}\,\Xi_{a} \Xi_{b}} \, \left(r_{h}^{\,2}+a^{2}\right) \, \left(r_{h}^{\,2}+b^{2}\right) ~.
\end{equation}
and the Hawking temperature $T=\beta^{-1}$ is given by
\begin{equation}\label{HTKerrAdS5}
  T = \frac{r_{h}}{2\pi} \, \left(1+\frac{r_{h}^{\,2}}{\ell^{2}}\right) \, \left(\frac{1}{r_{h}^{\,2}+a^{2}}
                 + \frac{1}{r_{h}^{\,2}+b^{2}}\right) - \frac{1}{2 \pi r_{h}} ~.
\end{equation}
Note that the metric used here has the same form as \eqref{CoordSystem}.

%%%%%%%%%%%%%%%%%%%%%%%%%%%%%%%%%%%%%%%%%%%%%%%%%%%%%%%%%%%%%
\subsection{Background Subtraction}
\label{sec:BSCalculation}
%%%%%%%%%%%%%%%%%%%%%%%%%%%%%%%%%%%%%%%%%%%%%%%%%%%%%%%%%%%%%

The renormalized action for Kerr-$AdS_{5}$ was calculated in \cite{Gibbons:2004ai} using background subtraction. The authors obtained
\begin{equation}\label{BSGamma}
   \Gamma = \frac{2 \pi^2 \beta}{\kappa^2 \, \ell^{2} \, \Xi_{a} \,\Xi_{b} }\,\left(m\ell^{2} - 
   \left(r_{h}^{\,2}+a^{2}\right)\left(r_{h}^{\,2}+b^{2}\right)\raisebox{12pt}{\,}\right) ~.
\end{equation}
The Quantum Statistical Relation (QSR) says that this renormalized action should be expressed in terms of the mass and other conserved charges of Kerr-$AdS_{5}$ according to
\begin{equation}\label{QSR}
  \beta^{-1} \, \Gamma = M - T\,S-\Omega_{a} J_{a} - \Omega_{b} J_{b}
\end{equation}
where $S$ is the entropy, and the $J_{i}$ are the angular momenta conjugate to the angular velocities
\eqref{AVKerrAdS5}. In terms of these quantities, the mass $M$ should satisfy the first law of black hole thermodynamics
\begin{equation}\label{FirstLaw}
  dM = T \,dS + \Omega_{a} dJ_{a}+ \Omega_{b} dJ_{b} ~.
\end{equation}
Thus, we identify $\beta^{-1} \Gamma$ with the thermodynamic potential $\Phi$ for the system, and can calculate the entropy and angular momenta according to the standard thermodynamic identities
\begin{eqnarray} \label{ThermoIdentities}
  S & = & - \left. \frac{\partial \Phi}{\partial T} \right|_{\Omega_{i}} \\ \nonumber
  J_{i} & = & - \left. \frac{\partial \Phi}{\partial \Omega_{i}} \right|_{T, \Omega_{j \neq i}} ~.
\end{eqnarray}
Using the expressions for the angular velocities \eqref{AVKerrAdS5} and Hawking Temperature \eqref{HTKerrAdS5}, one finds
\begin{eqnarray}\label{BSCharges} \nonumber
   S & =& \frac{4 \,\pi^{3}}{\kappa^2 \,r_{h}\,\Xi_{a} \Xi_{b}} \, \left(r_{h}^{\,2}+a^{2}\right) \, \left(r_{h}^{\,2}+b^{2}\right) \\   
   J_{a} &=& \frac{4\, \pi^2 \, m \,a}{\kappa^2\,\Xi_{a}^{\,2}\,\Xi_{b}} \\ \nonumber
   J_{b} &=& \frac{4\, \pi^2 \, m \,b}{\kappa^2 \,\Xi_{b}^{\,2}\,\Xi_{a}} ~.
\end{eqnarray} 
As expected, the entropy is given by one quarter of the area in Planck units.  Although this result is often
assumed in the literature, it is easy to verify it directly using \eqref{ThermoIdentities}. Having obtained the entropy and the angular momenta one can verify that the quantity
\begin{equation}\label{TotalDifferential}
   T \, dS + \Omega_{a}\,dJ_{a} + \Omega_{b}\,dJ_{b}
\end{equation}
which appears in \eqref{FirstLaw} is indeed a total differential. Integrating \eqref{TotalDifferential} gives the mass of the spacetime
\begin{equation}
\label{BackgroundSubtractionMass}
   M =  \frac{2 \pi^{2} m}{ \kappa^2 \,\Xi_{a}^{\,2}\,\Xi_{b}^{\,2}}\,\left(2 \,\Xi_{a} +
            2\,\Xi_{b}-\Xi_{a}\,\Xi_{b}\right) + M_{0} ~.
\end{equation}
The last term, $M_{0}$, is an integration constant which should not depend on $S$, $J_{a}$, or $J_{b}$, or alternately on the parameters $m$, $a$, and $b$. Typically $M_{0}$ is set to zero, which gives a mass that satisfies the QSR \cite{Gibbons:2004ai} and agrees with results obtained using a wide range of techniques \cite{Das:2000cu,Deser:2005jf,Deruelle:2004mv,Hollands:2005wt,Hollands:2005ya}. More importantly, it is consistent with a vanishing mass for global $AdS$ \cite{Ashtekar:1999jx}.

%%%%%%%%%%%%%%%%%%%%%%%%%%%%%%%%%%%%%%%%%%%%%%%%%%%%%%%%%%%%%
\subsection{Boundary Counterterms}
%%%%%%%%%%%%%%%%%%%%%%%%%%%%%%%%%%%%%%%%%%%%%%%%%%%%%%%%%%%%%

We now want to calculate the renormalized on-shell action using the boundary counterterm method. This calculation was first performed using the minimal subtraction scheme in \cite{Awad:1999xx}, and has subsequently been reviewed in several references \cite{Awad:2000ac,Awad:2000aj,Awad:2000ie}. Because the minimal subtraction calculation disagrees with the results of the previous section we will examine this approach in detail. Our renormalized action is
\begin{equation}
\label{BCGamma}
    \Gamma_{HJ} = I - I_{HJ}
\end{equation}
where $I$ is given in \eqref{GravityAction}. The counterterm action $I_{HJ}$ consists of the minimal subtraction terms \eqref{MSCTaction} and the finite term $I_{CT}^{(4)}$ which is needed to enforce bulk diffeomorphism invariance of $\Gamma_{HJ}$. While it is relatively simple to calculate $I$ and $I_{MS}$, evaluating  $I_{CT}^{(4)}$ is more involved.

%%%%%%%%%%%%%%%%%%%%%%%%%%%%%%%%%%%%%%%%%%%%%%%%%%%%%%%%%%%%%
\subsubsection{The Einstein-Hilbert and Gibbons-Hawking Terms}
%%%%%%%%%%%%%%%%%%%%%%%%%%%%%%%%%%%%%%%%%%%%%%%%%%%%%%%%%%%%%

The first term in the action consists of the Einstein-Hilbert and Gibbons-Hawking-York terms
 \begin{equation}
   I_{EH} +I_{GHY}  =  -\frac{1}{2\,\kappa^{2}}\,\int_{\MM_c}\nts\nts d^{5}x \,\sqrt{g} \, \left(  R - 2 
       \Lambda \right) - \frac{1}{\kappa^{2}}\,\int_{\Sigma_c} \nts \nts d^{4}x \, \sqrt{\gamma}\,K ~.
 \end{equation}
The Einstein-Hilbert term is simple to evaluate, and gives
\begin{equation}
  I_{EH} = \frac{2 \pi^{2} \beta}{\kappa^{2}\,\ell^{2}\,\Xi_{a}\Xi_{b}} \, \left( r_{c}^{4} + r_{c}^{2}\,(a^{2}+b^{2})
      -(r_{h}^{\,2}+a^{2})\,(r_{h}^{\,2}+b^{2})+a^{2}\,b^{2} \raisebox{12pt}{\,}\right) ~.
\end{equation}
From the metric \eqref{KerrAdS5Metric} we can identify the unit normal vector $n^{\mu}$ and construct the trace of the extrinsic curvature, K
\begin{equation}
   K = \frac{1}{\sqrt{g}}\,\partial_{r}\left(\sqrt{g}\,\sqrt{\frac{V-2m}{U}}\right) ~.
\end{equation}
With this expression, the Gibbons-Hawking-York term evaluates to
\begin{eqnarray}
  I_{GHY}  & = &  \frac{2\pi^{2}\beta}{\kappa^{2}\,\ell^{2}\,\Xi_{a}\Xi_{b}}\,\left(-4\,r_{c}^{4}-\frac{1}{2}\,
      \left(5\,(a^{2}+b^{2})+6\,\ell^{2}\right) \,r_{c}^{2} \right. \\ \nonumber
      & & \hspace{1cm}\left.+ \frac{1}{6}\,\ell^4 \, \left(\Xi_{a}-\Xi_{b}\right)^{2} - a^{2}\,b^{2}
            -\frac{3}{2}\,\ell^{2}\,(a^{2}+b^{2}) + 4 \,m\,\ell^{2}\, \right) ~.
\end{eqnarray}
Together, $I_{EH}$ and $I_{GHY}$ give the regulated on-shell action 
\begin{eqnarray}\label{RegulatedAction}
  I & = &  \frac{2 \pi^{2} \beta}{\kappa^{2}\,\ell^{2}\,\Xi_{a}\Xi_{b}}\,\left( -3\,r_{c}^{4}
   -\frac{3}{2}\,(a^{2}+b^{2})\,r_{c}^{2}-3 \,\ell^{2} \,r_{c}^{2}
   -(r_{h}^{\,2}+a^{2})\,(r_{h}^{\,2}+b^{2})\right. \\ \nonumber
    & & \hspace{1cm} \left. \frac{1}{6}\,\ell^4 \, \left(\Xi_{a}-\Xi_{b}\right)^{2}-\frac{3}{2}\,\ell^{2}\,(a^{2}+b^{2})
     + 4 \,m\,\ell^{2}\,\right) ~.
\end{eqnarray}

%%%%%%%%%%%%%%%%%%%%%%%%%%%%%%%%%%%%%%%%%%%%%%%%%%%%%%%%%%%%%
\subsubsection{The Counterterm Action}
%%%%%%%%%%%%%%%%%%%%%%%%%%%%%%%%%%%%%%%%%%%%%%%%%%%%%%%%%%%%%

 Next we evaluate the minimal subtraction counterterms
 \begin{equation}
   I_{MS} = - \frac{1}{\kappa^{2}}\,\int_{\dM} \bns d^{4}x \sqrt{\gamma}\,\left(\frac{3}{\ell}
            + \frac{\ell}{4}\,\RR \right) ~.
 \end{equation}
There are no subtleties in evaluating these terms, which give
\begin{eqnarray}
  I_{MS} & = & \frac{2\pi^{2}\beta}{\kappa^{2}\,\ell^{2}\,\Xi_{a}\Xi_{b}}\,\left( -3\,r_{c}^{4} 
      -\frac{3}{2}\,(a^{2}+b^{2})\,r_{c}^{2} - 3 \ell^{2}\,r_{c}^{2} + \frac{1}{8}\,\ell^4 \, 
      \left(\Xi_{a}-\Xi_{b}\right)^{2}\right. \\ \nonumber
      & & \hspace{1cm} \left. -\frac{3}{8}\,\ell^{4}\,\Xi_{a}\,\Xi_{b} - \frac{3}{2}\,\ell^{2}(a^{2}+b^{2}) 
      -3 \,m\,\ell^{2}\right) ~.
\end{eqnarray}
These counterterms are sufficient to remove the infra-red divergent terms from the regulated on-shell action.
Combining them with the terms in \eqref{RegulatedAction} we obtain the minimal subtraction renormalized action
\begin{equation}\label{IminusICT}
  \Gamma_{MS} = \frac{2\pi^{2}\beta}{\kappa^{2}\,\ell^{2}\,\Xi_{a}\Xi_{b}}\,\left( m\,\ell^{2}
       -(r_{h}^{\,2}+a^{2})\,(r_{h}^{\,2}+b^{2}) + \frac{3}{8}\,\ell^{4}\,\Xi_{a}\,\Xi_{b} 
       + \frac{\,\ell^4}{24}\,\left(\Xi_{a}-\Xi_{b}\right)^{2} \right) ~.
\end{equation}
This is the same action found in \cite{Awad:1999xx}. If one were to treat this quantity as the full renormalized action and use the entropy and angular momenta derived in the previous section~\footnote{Using \eqref{IminusICT} in \eqref{ThermoIdentities} does not yield \eqref{BSCharges}!},
then the quantum statistical relation would imply that the total mass is given by
\begin{equation}\label{Mprime}
   M_{MS} = M + \frac{2 \pi^{2}}{\kappa^{2}}\, \left( \frac{3}{8}\,\ell^{2} + \frac{\,\ell^{2}}{24
}\,\frac{(\Xi_{a}-\Xi_{b})^{2}}{\Xi_{a}\,\Xi_{b}}\right) ~.
\end{equation}
This expression for the mass differs from \eqref{BackgroundSubtractionMass} by two terms which represent the
Casimir energy associated with a dual conformal field theory living in the rotating Einstein static universe \cite{Hawking:1998kw, Cai:2005kw, Gibbons:2005vp, Gibbons:2005jd}

For certain values of the rotation parameters $a$ and $b$ the difference between $M$ and $M_{MS}$ is fairly innocuous. For instance, if we consider the non-rotating black hole, $a=b=0$, then the Casimir energy is a constant shift to $M$ which depends only on the cosmological constant
\begin{equation}
  M_{Cas} = \frac{2 \pi^{2}}{\kappa^{2}}\, \frac{3}{8}\,\ell^{2}  ~.
\end{equation}
From the point of view of the first law this can be regarded as the integration constant $M_{0}$ that appeared in \eqref{BackgroundSubtractionMass}. Because this Casimir term does not depend on the mass parameter $m$ it does not affect the differential form of the first law, which is still satisfied. The same is true when both rotation parameters are equal, $a=b$.

For the general rotating black hole the Casimir shift in \eqref{Mprime} is not harmless. In this case it is given by
\begin{equation}
  M_{Cas} = \frac{2 \pi^{2}}{\kappa^{2}}\,  \left( \frac{3}{8}\,\ell^{2} + \frac{\,\ell^{2}}{24} \,\frac{(\Xi_{a}-\Xi_{b})^{2}}{\Xi_{a}\,\Xi_{b}}\right) ~.
\end{equation}
Repeating the thermodynamic analysis of the previous section with the action \eqref{IminusICT}, one finds that the first law is \emph{not} satisfied, as pointed out in \cite{Gibbons:2004ai}.

In the next section we will calculate the remaining part of the counterterm action, and show that the resulting renormalized action is consistent with the first law. We should also remind the reader that
\cite{Papadimitriou:2005ii} presents another explanation for this apparent breakdown of the first law.

%%%%%%%%%%%%%%%%%%%%%%%%%%%%%%%%%%%%%%%%%%%%%%%%%%%%%%%%%%%%%
\subsubsection{The Finite Term \texorpdfstring{$I_{CT}^{(4)}$}{}}
\label{sec:FiniteTerm}
%%%%%%%%%%%%%%%%%%%%%%%%%%%%%%%%%%%%%%%%%%%%%%%%%%%%%%%%%%%%%

We will now calculate the counterterm $I_{CT}^{(4)}$, given by
\begin{equation}\label{FiniteTerm}
 I_{CT}^{(4)}  =  -\frac{1}{\kappa^{2}}\,\lim_{d \to 4^{+}} \int_{\Sigma_c} \nts \nts d^{d}x \, \sqrt{\gamma} 
        \, \frac{\ell^{3}}{2(d-2)^{2}(d-4)}\,\left(\RR^{ab}\RR_{ab}-\frac{d}{4(d-1)}\,\RR^{2}\right) ~.
\end{equation}
This is a non-trivial calculation that requires an evaluation of the integrand away from $d=4$. This is complicated by the fact that certain quantities characterizing the metric, such as the number of free rotation parameters, are not continuous functions of $d$. However, we will find that there are simplifications that will allow us to obtain the integrand for arbitrary $d$. The term \eqref{FiniteTerm} will then precisely cancel the terms in the action \eqref{IminusICT} which interfere with the first law.

In section \ref{sec:FiniteCounterterm} we saw that \eqref{FiniteTerm} depends only on the conformal representative obtained from $\gamma_{ab}$ via the rescaling \eqref{InducedRepRelation}. Much of the analysis that follows is simplified by discarding the physical metric at this point and working directly with $\hat{h}_{ab}$. Using \eqref{GeneralKerrAdS}, the conformal representative is 
\begin{eqnarray}\label{KerrAdSConfRep}
  d\hat{s}^{2} & = &  -W \, dt^{2} + \ell^{2}\,\sum_{i=1}^{N}\,\frac{\mu_{i}^{\,2}}{\Xi_{i}}\,
    \left(d \varphi_{i}+ \frac{a_{i}}{\ell^{2}}\,dt\right)^{2} + \ell^{2}\,\sum_{i=1}^{N+\eps}
\frac{1}{\Xi_{i}}\,d\mu_{i}^{\,2} \\ \nonumber
       & & - \frac{\ell^{2}}{W}\,\left( \sum_{i=1}^{N+\eps} \, \frac{1}{\Xi_{i}}\,\mu_{i}\,d\mu_{i}\right)^{2}
\end{eqnarray}
where $N=[d/2]$ is the integer part of $d/2$, and $\eps=d-2N$. Now we make the following observations, which will allow us to calculate \eqref{FiniteTerm} directly.
\begin{enumerate}
  \item Because we want the $d \to 4^+$ limit, we treat $d$ as if it were an even number, so that $\eps=0$ and $N = d/2$.
  \item The number of rotation parameters is not a continuous function of $d$. But all Kerr-$AdS$ metrics with $d+1 \geq 5$ have at least two such parameters. We can think of these metrics as members of a family of metrics with two non-zero rotation parameters that are defined for $d \geq 4$. We can write down the general form of a metric in this family, and continue its form to non-integer $d$.
\end{enumerate}
For the purposes of this calculation we will actually simplify a bit further, and set all $a_{i}=0$ for $i \neq 1$. This implies that $\Xi_{i}=1$ for all $i \neq 1$. We will perform the calculation explicitly for this case, and then present the result of an identical calculation for \emph{two} non-zero rotation parameters. 

We begin the calculation by defining the following new coordinates
\begin{equation}
  \mu_{1} = \mu \hspace{.5cm} \varphi_{1} = \phi \hspace{.5cm} \varphi_{N} = \zeta
\end{equation}
\begin{equation}   
  \varphi_{i+1}=\psi_{i} \hspace{.25cm} \forall \, \,  i = 1,\ldots, N-2
\end{equation}
\begin{equation}
  \mu_{\alpha+1} = \nu^{\alpha} \hspace{.25cm} \forall \,\, \alpha = 1,\ldots, N-2 ~.
\end{equation}
In terms of these variables, the constraint \eqref{muConstraint} on the $\mu_{i}$ becomes
\begin{equation}
   \mu_{N}^{\,2} = 1 - F(\mu,\nu^{\alpha})
\end{equation}
where the function $F(\mu,\nu^{\alpha})$ is given by
\begin{equation}
  F(\mu,\nu^{\alpha}) = \mu^{2} + \sum_{\alpha=1}^{N-2} (\nu^{\alpha})^{2} ~.
\end{equation}
We are now able to write the metric \eqref{KerrAdSConfRep} in the following form
\begin{eqnarray} \label{ConformalMetric}
 d\hat{s}^{2} & = & - dt^{2} + 2\,\frac{a}{\Xi_{a}}\,\mu^{2}\,dt\,d\phi + \ell^{2}\,\frac{\mu}{\Xi_{a}}\,d\phi^{2} + 
     \ell^{2}\,(1-F(\mu,\nu^{\alpha}))\,d\zeta^{2} \\ \nonumber
 & &  + \, \ell^{2}\,\sum_{i}^{N-2}(\nu^{i})^{2}\,d\psi_{i}^{2} + \ell^{2}\,\left(
   \frac{1}{\Xi_{a}}+\frac{\mu^{2}}{1-F(\mu,\nu^{\alpha})} 
 - \frac{\mu^{2}}{W}\,\frac{a^{4}}{\ell^{4}\,W^{2}} \right) d\mu^{2} \\ \nonumber
 & & + 2\,\ell^{2}\,\sum_{\alpha=1}^{N-2}\frac{\mu\,\nu^{\alpha}}{1-F(\mu,\nu^{\alpha})}\,d\mu\,d\nu^{\alpha}
 + \, \ell^{2}\sum_{\alpha=1}^{N-2}\sum_{\beta=1}^{N-2}\,\left(\delta_{\alpha\beta}+
\frac{\nu^{\alpha}\,\nu^{\beta}}{1-F(\mu,\nu^{\alpha})}\right) d\nu^{\alpha} d\nu^{\beta} ~.
\end{eqnarray}
This metric can be decomposed into four blocks along the diagonal. The first two are $2\times2$ and $1 \times
1$ blocks. The third block involves the $N-2$ coordinates $\psi^{i}$ and is diagonal, while the fourth block is $N-1 \times N-1$ and is in ADM form. Because of the relatively simple form of this metric it is straightforward to invert it and then calculate the curvature tensors. We obtain the following expressions for the scalar curvature and the square of the Ricci tensor
\begin{eqnarray}
\label{UnphysicalCurvatures}
  \hat{\RR} & = & \frac{(d-1)}{\ell^{4}}\,\left( (d-2)\,\ell^{2} - 4 a^{2}+ (d+2)\,a^{2}\,\mu^{2}\right) \\ \nonumber
  \hat{\RR}^{ab}\hat{\RR}_{ab} & = & \frac{(d-1)(d-2)^{2}}{\ell^{4}} + 4\frac{a^{2}}{\ell^{6}}\,(d-2)\,
         \left( \left(\frac{1}{2}\,d^{2}-1\right)\mu^{2}-(2d-3)\right) \\ \nonumber
         & & + 4 \frac{a^{4}}{\ell^{8}}\,\left(\frac{1}{2}\,(d^{2}+2d-4) - \frac{1}{2}\,d\,(5d-6)\,\mu^{2}
             + \frac{1}{4}\,(d-1)\,(d^{2}+4d-4)\,\mu^{4}\right) ~.
\end{eqnarray}
Equipped with these expressions, we are now able to calculate the final piece of the counterterm
action.

Using \eqref{UnphysicalCurvatures} we can write the integrand of $I_{CT}^{(4)}$ as a polynomial that is quadratic
in $\mu^2$
\begin{equation}\label{mupolynomial}
 \frac{\ell^{3}}{2(d-2)^{2}(d-4)}\,\left(\hat{\RR}^{ab}\hat{\RR}_{ab}-\frac{d}{4(d-1)}\,\hat{R}^{2}\right)
    = b_{0} + b_{1}\,\mu^{2} + b_{2}\,\mu^{4} ~.
\end{equation}
Thus, we need to evaluate integrals of the form
\begin{equation}\label{WeNeedThese}
  \int d\mu \, d^{N-2}\nu^{\alpha}\,\sqrt{\hat{h}}\,\mu^{2n}
\end{equation}
where the square-root of the determinant of the conformal representative is given by
\begin{equation}\label{UnphysicalVolume}
   \sqrt{\hat{h}} = \ell^{d-1}\,\frac{\mu}{\Xi_{a}}\,\prod_{\alpha=1}^{N-2}\,\nu^{\alpha} ~.
\end{equation}
Evaluating the integral \eqref{WeNeedThese} gives
\begin{equation}\label{RandomIntegral}
  \int d\mu \, d^{N-2}\nu^{\alpha}\,\sqrt{\hat{h}}\,\mu^{2n} = \frac{2^{N-2}\,\ell^{d-1}}{\Xi_{a}} \, 
       \left( \prod_{k=1}^{N-1} (2k+2n)\right)^{-1} ~.
\end{equation}
By shifting the range of the product appearing in \eqref{RandomIntegral} it is easy to show that
\begin{equation}
  \int d\mu \, d^{N-2}\nu^{\alpha}\,\sqrt{\hat{h}}\,\mu^{2n} =   
     \frac{n! \,(N-1)!}{(N-1+n)!}\,\int d\mu \, d^{N-2}\nu^{\alpha}\,\sqrt{\hat{h}} ~.
\end{equation}
This result allows us to rewrite the integral of \eqref{mupolynomial} in terms of a boundary volume integral
\begin{equation}
  I_{CT}^{(4)}  = \frac{1}{8 \pi G} \, \lim_{d \to 4^+} \left(b_{0} + b_{1}\,\frac{2}{d} 
    + b_{2}\, \frac{8}{d(d+2)}  \right) \, \int_{\dM}\bns d^{d}x \, \sqrt{\hat{h}} ~.
\end{equation}
The coefficients $b_{i}$ are given by
\begin{eqnarray}
b_{0} & = & - \frac{(d-1)}{8 \ell} + \frac{(d-3)}{(d-4)}\,\frac{a^{2}}{\ell^{3}} -
\frac{1}{(d-4)}\,\frac{a^{4}}{\ell^{5}} \\ \nonumber
b_{1} & = & - \frac{d^{2}-d-4}{4(d-4)}\,\frac{a^{2}}{\ell^{3}} + \frac{d}{d-4}\,\frac{a^{4}}{\ell^{5}} \\ \nonumber
b_{2} & = & - \frac{(d-1)(d+4)}{8(d-4)}\, \frac{a^{4}}{\ell^{5}} ~.
\end{eqnarray}
With these expressions we finally obtain
\begin{equation}
  \left. I_{CT}^{(4)} \, \right|_{b=0} = \frac{2\pi^{2} \beta}{\kappa^2\,\ell^{2}\,\Xi_{a}}\,\left(
   \,\frac{3}{8}\,\ell^{4}\,\Xi_{a} + \frac{1}{24}\,a^{4}\right) ~.
\end{equation}
In the $b=0$ case this is precisely the difference between the action obtained using minimal subtraction \eqref{IminusICT}, and the action obtained using background subtraction \eqref{BSGamma}.

We can repeat the analysis in this section with both $a$ and $b$ non-zero. This does not introduce any new conceptual difficulties; it only makes the book-keeping aspects more tedious. We will simply quote the final
result of this calculation, which is
\begin{equation}\label{FullFiniteCounterterm}
  I_{CT}^{(4)} = \frac{2\pi^{2} \beta}{\kappa^2\,\ell^{2}\,\Xi_{a}\,\Xi_{b}}\,\left(
    \, \frac{3}{8}\,\ell^{4}\,\Xi_{a}\,\Xi_{b} + \frac{\,\ell^4}{24} \, (\Xi_{a}-\Xi_{b})^{2}\right) ~.
\end{equation}
Combining this with the other terms in the renormalized action, we obtain
\begin{equation}\label{FinalAction}
  \Gamma_{HJ} =  \frac{2 \pi^{2} \beta}{\kappa^2\,\ell^{2} \, \Xi_{a}\,\Xi_{b}} \, 
      \left(  m \, \ell^{2} - (r_{h}^{\,2}+a^{2})\,(r_{h}^{\,2}+b^{2}) \raisebox{12pt}{\,}\right)
\end{equation}
This is the same result that is found using background subtraction.

%%%%%%%%%%%%%%%%%%%%%%%%%%%%%%%%%%%%%%%%%%%%%%%%%%%%%%%%%%%%%
\subsection{A Comment on the Ambiguity of Balasubramanian and Kraus}
\label{sec:Ambiguity}
%%%%%%%%%%%%%%%%%%%%%%%%%%%%%%%%%%%%%%%%%%%%%%%%%%%%%%%%%%%%%

The authors of \cite{Balasubramanian:1999re} identify a natural ambiguity in the definition of the counterterm action for asymptotically $AdS_5$ spacetimes. It is instructive to ask whether this encompasses the finite term that was calculated in the previous section. Specifically, Balasubramanian and Kraus point out that one may supplement the counterterm action with terms that are quadratic in the boundary curvatures. Although they adopt a basis for the finite terms similar to the one we use in section \ref{sec:FiniteCounterterm}, for our purposes it is sufficient to consider a contribution to the counterterm action of the form
\begin{equation}\label{BKextra}
  \Delta I_{CT} = - \frac{\, \ell^3}{\kappa^2}\,\int_{\dM} \bns d^{4}x \sqrt{\hat{h}}  \, \left( c_1 \, \hat{\RR}^2 
   + c_2 \, \hat{\RR}^{ab}\hat{\RR}_{ab} + c_3 \, \hat{\RR}^{a}_{\,\,\,bcd}\hat{\RR}_{a}^{\,\,\,bcd}\right)
\end{equation}
where the $c_i$ are \emph{finite} numerical coefficients. Evaluating this expression for the Kerr-$AdS_5$ solution gives
\begin{equation}\label{BkextraEvaluated}
  \Delta I_{CT} \sim \frac{2\,\pi^2\,\beta}{\kappa^2\,\ell^2 \, \Xi_a \, \Xi_b}\,\left(3 c_1 + c_2 + c_3 \right) \, 
   \left(\ell^4\,\Xi_a \, \Xi_b + \ell^4 \left(\Xi_a - \Xi_b\right)^2 \raisebox{12pt}{\,}\right) ~.
\end{equation}
It is clear that no choice of the coefficients in \eqref{BKextra} can reproduce the result \eqref{FullFiniteCounterterm} from the previous section. 

The difference between the ambiguity pointed out in \cite{Balasubramanian:1999re} and the term $I_{CT}^{(4)}$ can be understood as the difference between trivial and non-trivial cocycles of the Weyl group in four dimensions \footnote{The abelian group of local conformal transformations of the metric.}, as described in \cite{Mazur:2001aa, Imbimbo:1999bj, Schwimmer:2000cu, Schwimmer:2003eq}. Recall from section \ref{sec:CountertermsAndDiffeos} that diffeomorphism invariance is broken in the minimal subtraction scheme by the terms in the field theory effective action that are responsible for the conformal anomaly 
\begin{equation}
  \delta_{\eps} \Gamma_{MS} \sim \int_{\dM}\bns d^{2k}x \sqrt{\hat{h}}\,\AA[\hat{h}]\,n_{\mu}\,\eps^{\mu} ~.
\end{equation}
The conformal anomaly is due to terms in the action which are non-local functionals of the conformal representative $\hat{h}_{ab}$ in the physical dimension $d=2k$. These terms may, however, be obtained from the $d \to 2k$ limit of functionals of the boundary metric which are perfectly local away from $d=2k$. The authors of \cite{Mazur:2001aa} show that dimensional regularization provides a simple and direct away of obtaining these terms, which generate non-trivial cocycles of the Weyl group in $d=2k$. The counterterm $I_{CT}^{(2k)}$ identifies and removes the non-trivial cocycles of the Weyl group in the gravitational action that are responsible for the breakdown in diffeomorphism invariance or, equivalently, the terms in the field theory effective action responsible for the conformal anomaly. The local terms in \eqref{BKextra}, on the other hand, correspond to trivial cocycles of the Weyl group in $d=4$, and cannot cancel the terms in the effective action that are generating the conformal anomaly~\footnote{It should be noted that, given our definition of an asymptotically $AdS$ spacetime, the only conformal anomaly present in the dual theory is type A, according to the classification of Deser and Schwimmer \cite{Deser:1993yx}.}. Comparing \eqref{BkextraEvaluated} with \eqref{FullFiniteCounterterm} shows this explicitly.

The techniques used in \cite{Mazur:2001aa} suggest an alternate means of calculating the counterterm $I_{CT}^{(4)}$ that does not involve dimensional continuation of the conformal representative $\hat{h}_{ab}$. This approach will be discussed in a future work \cite{McNees:2005xx}.

%%%%%%%%%%%%%%%%%%%%%%%%%%%%%%%%%%%%%%%%%%%%%%%%%%%%%%%%%%%%%
\subsection{Counterterm Charges}
\label{sec:CountertermCharges}
%%%%%%%%%%%%%%%%%%%%%%%%%%%%%%%%%%%%%%%%%%%%%%%%%%%%%%%%%%%%%

In \cite{Balasubramanian:1999re} the boundary counterterm method was used to renormalize the calculation of conserved charges via the quasi-local stress tensor of Brown and York \cite{Brown:1992br}. In this section we show that the stress tensor obtained from our modified action leads to the same conserved charges found in section \ref{sec:BSCalculation}.

The boundary stress tensor is defined by the functional derivative of the renormalized action with respect to
the boundary metric
\begin{eqnarray}
   T^{ab} & = & \frac{2}{\sqrt{\gamma}}\,\frac{\delta \, \Gamma}{\delta \, \gamma_{ab}} \\ \nonumber
          & = & 2 \, \left( \pi^{ab} - P^{ab}\right) ~.
\end{eqnarray}
We can use this stress tensor to calculate the conserved charges associated with various killing vectors of the boundary metric, which we write in the ADM form
\begin{equation}
  \gamma_{ab}\,dx^{a}\,dx^{b} = -N_{\Psi}\,dt^{2} + q_{IJ}\,\left(dx^{I} + N_{\Psi}^{\,I}dt \right)
      \,\left(dx^{J} + N_{\Psi}^{\,J}dt \right) ~.
\end{equation} 
In this decomposition $\Psi$ is a $d-1$ dimensional spacelike hypersurface, orthogonal to a timelike Killing
vector of unit norm which we denote $u^{a}$. If $\xi^{a}$ is an appropriately normalized killing
vector, then its associated conserved charge is given by
\begin{equation}
  Q_{\xi} = \int_{\Psi}\nts d^{d-1}x \sqrt{q}\,u^{a}\,T_{ab}\,\xi^{b} ~.
\end{equation}
Using this expression we can calculate the mass and angular momenta for the Kerr-$AdS_{5}$ spacetime, which are given by
\begin{eqnarray}
  M & = & \int_{\Psi} \nts d^{3}x \, \sqrt{q} \, N_{\Psi}\, u^{a}\,T_{ab}\,u^{b} \\
  J_{I} & = & \int_{\Psi} \nts d^{3}x \, \sqrt{q}\,q_{I}^{\,\,J}\,T_{a J}\,u^{a} ~.
\end{eqnarray}
To evaluate these integrals we first need to find the boundary stress tensor by taking the functional derivative of $\Gamma_{HJ}$ with respect to the boundary metric. For $d$ near the physical value $d=4$ we find
\begin{eqnarray}
  \kappa^2 \, T^{ab} & = & K^{ab} - \gamma^{ab}\,K - \frac{d-1}{\ell}\,\gamma^{ab} + \frac{\ell}{d-2}\,\left(\RR^{ab} 
    - \frac{1}{2}\,\gamma^{ab}\,\RR \right) \\ \nonumber
 & & + \frac{\ell^{3}}{(d-2)^{2}(d-4)}\,\FF^{ab} ~.
\end{eqnarray}
The tensor $\FF^{ab}$ contains the contributions from the counterterm $I_{CT}^{(4)}$, and is given
by
\begin{eqnarray}
   \FF^{ab} & = & -\frac{1}{2}\,\gamma^{ab}\,\left(\RR^{cd}\,\RR_{cd} -\frac{d}{4(d-1)}\,\RR^2 \right)  
           - \frac{d}{2\,(d-1)}\,\RR \, \RR^{ab} \\ \nonumber
      &  &  - \frac{1}{2\,(d-1)}\,\gamma^{ab}\,\DD^{2}\,\RR + \DD^{2}\,\RR^{ab}  
          - \frac{(d-2)}{2\,(d-1)}\,\DD^{a}\,\DD^{b}\,\RR  \\ \nonumber
      &  & + \, 2 \,\RR^{acbd} \,\RR_{cd} \raisebox{15pt}{\,}~.
\end{eqnarray}
Once again, we are forced to evaluate contributions from the new counterterm by continuing away from $d=4$.
Conceptually this calculation is no more difficult than the one in the previous section (Note that
the first two terms in $\FF^{ab}$ are proportional to $\gamma^{ab} \, \LL_{CT}^{(4)}$). As a practical
matter, however, it is much more tedious. We simply quote the results of these calculations, which give the same values predicted by the thermodynamic analysis
\begin{eqnarray}
  M & = &  \frac{2 \,\pi^{2}\,m}{\kappa^2 \, \Xi_{a}^{\,2} \Xi_{b}^{\,2}}\,\left( 2\,\Xi_{a} + 2\,\Xi_{b}-
\Xi_{a}\,\Xi_{b}\right) \\
  J_{a} & = & \frac{4\,\pi^2 \, m \, a}{\kappa^2\,\Xi_{a}^{\,2}\,\Xi_{b}}\\
  J_{a} & = & \frac{4\,\pi^2 \, m \, b}{\kappa^2\,\Xi_{a}\,\Xi_{b}^{\,2}} ~.
\end{eqnarray}
These results also agree with the conserved charges found in \cite{Das:2000cu,Deser:2005jf,Deruelle:2004mv,Hollands:2005wt,Hollands:2005ya}.

In conclusion we find that a boundary counterterm renormalization of the on-shell action that is consistent with the symmetries of the bulk gravitational theory leads to conserved charges that satisfy the first law of black hole thermodynamics and the quantum statistical relation. Furthermore, it is easy to see that these conserved charges vanish for global $AdS_5$, which corresponds to the $(m,a,b) = (0,0,0)$ case of the Kerr-$AdS$ metric. This should be sufficient to address the concerns of \cite{Ashtekar:1999jx}.

%%%%%%%%%%%%%%%%%%%%%%%%%%%%%%%%%%%%%%%%%%%%%%%%%%%%%%%%%%%%%
%%%%%%%%%%%%%%%%%%%%%%%%%%%%%%%%%%%%%%%%%%%%%%%%%%%%%%%%%%%%%
\section{Conclusion}
\label{sec:Conclusion}
%%%%%%%%%%%%%%%%%%%%%%%%%%%%%%%%%%%%%%%%%%%%%%%%%%%%%%%%%%%%%
%%%%%%%%%%%%%%%%%%%%%%%%%%%%%%%%%%%%%%%%%%%%%%%%%%%%%%%%%%%%%

In this paper we have shown that the boundary counterterm method \cite{Henningson:1998gx,Balasubramanian:1999re,Emparan:1999pm,deHaro:2000xn} can be modified to give a finite, diffeomorphism invariant action for asymptotically $AdS$ spacetimes. Normally, the counterterm procedure gives an action whose response to certain bulk diffeomorphisms is related to the conformal anomaly of the dual field theory \cite{Henningson:1998gx, deHaro:2000xn, Skenderis:2002wp, Papadimitriou:2005ii}. In section \ref{sec:CountertermHamiltonian} we obtained a new counterterm that restores full diffeomorphism invariance to the renormalized action. This construction is only necessary, of course, when the dimension of the boundary is even. Using our improved action to study the Kerr-$AdS_5$ black hole, we find conserved charges consistent with the first law of black hole thermodynamics. Our results also address the criticism of references like \cite{Ashtekar:1999jx}, since our action gives a vanishing mass for the pure $AdS_{2k+1}$.

The new counterterm is well defined for spacetimes which are asymptotically $AdS$, as defined in section \ref{sec:Conventions}. In that sense, our results are less complete than the very general treatment given in \cite{Papadimitriou:2005ii}. It would be interesting to investigate the extent to which our construction can be generalized. For instance, one might consider asymptotically $AdS$ spacetimes whose boundary topology is different than $\BR \times S^{d-1}$, but for which the Euler number is still zero. Some of the spacetimes considered in \cite{Emparan:1999pm} and \cite{deBoer:2004yu} fall into this category. On the other hand, it appears difficult to relax the condition that the conformal class of metrics induced on the boundary must be conformally flat. Another natural extension would be the inclusion of matter \cite{Skenderis:1999nb, deHaro:2000xn, Bianchi:2001de, Bianchi:2001kw, Skenderis:2002wp}, and the application of these techniques to black-hole spacetimes that arise in gauged supergravities \cite{Cvetic:2004ny,Chong:2004dy,Cvetic:2005zi,Kunduri:2005zg,Chong:2005da,Chong:2005hr,Chen:2005zj}. It would also be interesting, and presumably straightforward, to adapt the techniques described in this paper to the case of asymptotically de-Sitter spacetimes.

It is important to point out that our results do not represent a disagreement with the usual AdS/CFT treatment of the field theory dual of an asymptotically $AdS$ spacetime. It is clear that the counterterm construction given in \cite{Henningson:1998gx, Balasubramanian:1999re, Emparan:1999pm, Papadimitriou:2005ii} is the correct procedure for calculating the effective action of these theories. We have focused, instead, on questions one might ask of the bulk gravitational theory, independently of the $AdS$/CFT correspondence. In that case it is appropriate to use the diffeomorphism invariant action $\Gamma_{HJ}$, which differs from field theory effective action $\Gamma_{MS}$ when the boundary dimension is even
\begin{equation}
   \Gamma_{MS} = \Gamma_{HJ} + I_{CT}^{(2k)} ~.
\end{equation}
This last equation represents an interesting perspective, since it appears that the $d$-dimensional effective action of the dual field theory is related to a subtle breaking of $d+1$-dimensional diffeomorphism invariance in the gravitational theory.

Finally, any treatment of the boundary counterterm method is incomplete without raising the question of whether there is an analogous construction for the full, higher dimensional supergravities that appear in the $AdS$/CFT correspondence. As a tool in $AdS$/CFT, the counterterm approach is only relevant insofar as several features of the  dual field theory are captured by pure gravity with a negative cosmological constant. The universal properties of the counterterm construction seem to require the existence of a length scale as a natural parameter in the gravitational theory, but this is not usually the case in $AdS$/CFT. Rather, one encounters 10 and 11 dimensional supergravities that admit Freund-Rubin backgrounds, where a dynamical Ramond-Ramond field strength mimics a cosmological constant for some lower dimensional asymptotically $AdS$ spacetime. Thus, the $AdS$ length scale that appears in this case is a feature of a particular supergravity solution, and not of the supergravity theory itself \cite{Vijay:xxxx}. Still, there are examples where dynamical fields, and not fixed parameters, seem to provide a sufficient framework for the counterterm construction \cite{Larsen:2003pf, Larsen:2004kf, Papadimitriou:2004ap}. The treatment of divergences in the full supergravity action via the counterterm method remains an open and intriguing area of research.

\section{Acknowledgments}

The author would like to thank Vijay Balasubramanian, Ben Burrington, Josh Davis, Jason Kumar, Finn Larsen, Jim Liu, David Lowe, and Paul de Medeiros for useful discussions. In addition, special thanks go to Don Marolf for insightful comments and questions based on an early version of this paper, and Malcolm Perry for collaboration in the initial stages of this work. This work was supported by DoE grant DE-FG02-95ER40899 at the University of Michigan, and by DoE grant DE-FG02-91ER40688-Task A at Brown University. 

\newpage

\appendix

\section{Asymptotic Expansions}
\label{app:AsymptoticExpansions}

In section \ref{sec:GammaAndDiffeos} we claim that $\Gamma_{HJ}$ is invariant under diffeomorphisms
\begin{equation} \label{Claim}
  \delta_{\eps} \Gamma_{HJ} = - \int_{\dM} \bns d^{d}x \sqrt{\gamma}\,\left[ 2\,K_{ab}\,P^{ab} 
    + \frac{1}{\kappa^2}\, (\RR - 2 \Lambda) \right]\,n_{\mu}\,\eps^{\mu} = 0 ~.
\end{equation}
In this appendix we provide the details of this calculation. For simplicity's sake we will work with asymptotically $AdS_{5}$ spacetimes. As in section \ref{sec:CountertermHamiltonian} we use dimensional regularization, treating the boundary as if the dimension were $d=4+ 2\varepsilon$, with $\varepsilon$ small and positive. The method we use in this example extends to the general case of asymptotically $AdS_{2k+1}$ spacetimes, although the calculations can become very cumbersome.

We will work in Fefferman-Graham coordinates \cite{Graham:1985CI} where the metric takes the form
\begin{equation}
  ds^2 = \frac{\ell^2}{4\rho^2}\,d\rho^2 + \frac{1}{\rho}\,h_{ab}(x,\rho)\,dx^a\,dx^b ~.
\end{equation}
This choice of coordinate system allows us to make contact with a number of useful results in the literature regarding the asymptotic expansion of the metric and its curvature tensors \cite{Henningson:1998gx, Skenderis:1999nb, deHaro:2000xn, Bianchi:2001de, Bianchi:2001kw, Skenderis:2002wp}. As in section \ref{sec:Foliations} we foliate the region near $\rho=0$ with constant $\rho$ surfaces orthogonal to the spacelike unit normal vector
\begin{equation} 
     n_{\mu} = \frac{\ell}{2 \rho}\,\delta_{\mu \rho} ~.
\end{equation}
The induced metric on the constant $\rho$ surface is
\begin{equation}
   \gamma_{ab} = \frac{1}{\rho}\,h_{ab}(x,\rho) ~.
\end{equation}
We assume that $h_{ab}$ can be Taylor expanded around $\rho=0$ to give
\begin{equation}
  h_{ab}(x,\rho) = \hat{h}_{ab}(x) + \rho \, h_{ab}^{(1)}(x) + \rho^2 \, h_{ab}^{(2)}(x) + \dots ~.
\end{equation}
The leading term in this expansion is a representative of the conformal class of metrics on the surface $\rho=0$. The coefficients $h_{ab}^{(i)}$ are covariant functions of $\hat{h}_{ab}$ that are obtained by solving the Einstein equations \eqref{Einstein} order-by-order in $\rho$. An excellent discussion of this procedure can be found in, for instance, \cite{deHaro:2000xn}. To show \eqref{Claim} with $d=4 + 2\varepsilon$ we will only need a few terms from this expansion
\begin{eqnarray} \label{hab1}
  h_{ab}^{(1)} & = & - \frac{\ell^2}{(d-2)}\, \left( \hat{\RR}_{ab} - \frac{1}{2(d-1)}\,\hat{h}_{ab}
    \, \hat{\RR}\right) \\ \label{trace_hab2}
  \hat{h}^{ab}\,h_{ab}^{(2)}& = & \frac{\ell^4}{(d-2)^2}\,\left( \hat{\RR}^{ab}\,\hat{\RR}_{ab} - \frac{(3d-4)}{4(d-1)^2}\,\hat{\RR}^2\right) ~.
\end{eqnarray}
We will also need the following expansions for the intrinsic and extrinsic curvatures built from $\gamma_{ab}$, and the covariant volume element
\begin{eqnarray}
  \RR & = & \rho\,\hat{\RR} + \frac{1}{(d-2)}\,\rho^2 \, \left( \hat{\RR}^{ab}\hat{\RR}_{ab}
    - \frac{1}{2(d-1)}\,\hat{\RR}^2\right) + \OO(\rho^3) \\
  \RR_{ab} & = & \hat{\RR}_{ab} + \OO(\rho) \\
  K_{ab}   & = & -\frac{1}{\ell}\,\gamma_{ab} + \frac{1}{\ell}\,h_{ab}^{(1)} + \OO(\rho)  \\ \label{VolElement}
  \sqrt{\gamma} & = & \rho^{-\frac{d}{2}}\,\sqrt{\hat{h}}\,\left( 1 + \OO(\rho) \right) ~.
\end{eqnarray}
Now we can use these expressions to evaluate the integrand in \eqref{Claim}
\begin{equation}\label{NewExpression}
 \sqrt{\gamma}\,\left(2 K_{ab}\,P^{ab} + \frac{1}{\kappa^2}\,\left(\RR-2\Lambda\right)\right) = 
   \rho^{-2-\varepsilon}\,\sqrt{\hat{h}}\,(1+\OO(\rho))\,\left[-\frac{2}{\ell}\,P^{a}_{\,\,a} + \frac{2}{\ell}\,P^{ab}h_{ab}^{(1)}+\frac{1}{\kappa^2}\,\RR + \frac{d(d-1)}{\ell^2\kappa^2}\right] ~.
\end{equation}
Thus, proving \eqref{Claim} amounts to showing that the cut-off dependence of the quantity in square brackets is at most $\OO(\rho^3)$, so that the total integrand is $\OO(\rho^{1-\varepsilon})$ and vanishes as the cut-off is removed.

The counterterm momentum that appears in \eqref{NewExpression}, including the contribution from the counterterm $I_{CT}^{(4)}$, is
\begin{eqnarray}\label{CTMomentum2}
P^{ab} & = & \frac{(d-1)}{2\kappa^2 \ell}\,\gamma^{ab} - \frac{\ell}{2\kappa^2 (d-2)}\,\left(\RR^{ab} - 
   \frac{1}{2}\,\gamma^{ab}\RR\right) \\ \nonumber
& & + \frac{\ell^3}{4\kappa^2 d(d-2)^2}\,\gamma^{ab}\,\left(\RR^{ab}\RR_{ab} - \frac{d}{4(d-1)}\,\RR^2\right)
+ \ldots ~.
\end{eqnarray}
The `$\ldots$' represents a traceless term that is quadratic in the boundary curvature, but does not contribute to \eqref{NewExpression} at the orders in $\rho$ that we are interested in. We can now evaluate each of the terms in \eqref{NewExpression}. After replacing the first term using the trace of \eqref{CTMomentum2},  \eqref{NewExpression} becomes
\begin{equation}
  2 K_{ab} P^{ab} + \frac{1}{\kappa^2}\,\left(\RR-2\Lambda\right) = 
    \frac{2}{\ell}\,P^{ab}\,h_{ab}^{(1)} + \frac{1}{2\kappa^2}\,\RR - \frac{\ell^2}{2\kappa^2 (d-2)^2}\,\left(\RR^{ab}\RR_{ab} - \frac{d}{4(d-1)}\,\RR^2\right)
\end{equation}
Evaluating the remaining terms on the right hand side using \eqref{hab1} - \eqref{VolElement} shows that all terms proportional to $\rho$ or $\rho^2$ cancel, leaving only terms which are $\OO(\rho^3)$. Thus
\begin{equation}
\delta_{\eps} \Gamma_{HJ} = - \int_{\dM} \bns d^{4+2\varepsilon}x \sqrt{\hat{h}}\,\OO(\rho^{1-\varepsilon})\,n_{\mu}\eps^{\mu} ~.
\end{equation}
Although $n_{\mu} \sim \rho^{-1}$, the leading terms in $\epsilon^{\rho}$ for a diffeomorphism that is consistent with the asymptotics implied by \eqref{Einstein} are 
\begin{equation}
   \epsilon^{\rho} \sim  a(x) \, \rho + b(x)\, \rho \, \log{\rho}
\end{equation}
Therefore, the $\rho$ dependence of the leading order term in $n_{\mu}\eps^{\mu}$ is no greater than $\OO(\log{\rho})$, so that in the $\rho \to 0$ limit we find
\begin{equation}
\delta_{\epsilon} \Gamma_{HJ} = 0 ~.
\end{equation}
The action $\Gamma_{HJ}$ is invariant under diffeomorphisms, as we claimed at the end of section \ref{sec:GammaAndDiffeos}.

\section{The Kerr-\texorpdfstring{$AdS_{d+1}$}{AdS(d+1)} Solution}
\label{app:GeneralKerrAdS}
The general form of the Kerr-$AdS_{d+1}$ solution was obtained in \cite{Gibbons:2004uw,Gibbons:2004js}. For our purposes it is useful to express the solution in Boyer-Lindquist coordinates
\begin{eqnarray}\label{GeneralKerrAdS}
    ds^{2} & = & -W\,(1+\frac{r^{2}}{\ell^{2}})\,dt^{2} + \frac{U}{V-2m}\,dr^{2} + \frac{2m}{U}\,
                         \left(dt - \sum_{i=1}^{N} \frac{a_{i}\,\mu_{i}^{\,2}}{\Xi_{i} }\,d\vf_{i}\right)^{2} \\ \nonumber
                &    & + \sum_{i=1}^{N}\frac{r^{2}+a_{i}^{\,2}}{\Xi_{i}}\,\mu_{i}^{\,2}\,\left(d\vf_{i} + 
                               \frac{a_{i}}{\ell^{2}}\,dt\right)^{2}  + 
                               \sum_{i=1}^{N+\eps}\frac{r^{2}+a_{i}^{\,2}}{\Xi_{i}}\,d\mu_{i}^{\,2} \\
                               \nonumber
                &    & -\frac{1}{\ell^{2} W}\,\left(1+\frac{r^{2}}{\ell^{2}}\right)^{-1}\,\left( \sum_{i=1}^{N+\eps}
                               \frac{r^{2}+a_{i}^{\,2}}{\Xi_{i}} \mu_{i}\,d\mu_{i}\right)^{2}     ~.     
 \end{eqnarray}
In $d+1$ dimensions there are $N=[d/2]$ azimuthal angles $\vf_{i}$, with corresponding rotation parameters $a_{i}$,  and
 $N+\eps$ direction cosines $\mu_{i}$, with $\eps = d-2N$. The direction cosines $\mu_{i}$ satisfy the
constraint
\begin{equation}\label{muConstraint}
  \sum_{i}^{N+\eps} \mu_{i}^{\,2} = 1 ~.
\end{equation}
The functions $W$, $V$,  $U$, and $\Xi_{i}$ are given by
\begin{eqnarray}\label{KerrXi}
  \Xi_{i} & = & 1-\frac{a_{i}^{\,2}}{\ell^{2}} \\ \label{KerrW} 
   W & = & \sum_{i=1}^{N+\eps}\frac{\mu_{i}^{\,2}}{\Xi_{i}} \\ \label{KerrU}
   U & = & r^{\eps}\,\sum_{i=1}^{N+\eps}\frac{\mu_{i}^{\,2}}{r^{2}+a_{i}^{\,2}}\,\prod_{j=1}^{N} 
                \left(r^{2}+a_{j}^{\,2}\right) \\ \label{KerrV}
   V & = & r^{\eps-2}\,\left(1+\frac{r^{2}}{\ell^{2}}\right)\,\prod_{i=1}^{N}\left(r^{2}+a_{i}^{\,2}\right) ~.
\end{eqnarray}
When the dimension $d+1$ of spacetime is even there are $N+1$ direction cosines, but there are only $N$ azimuthal
angles and $N$ free rotation parameters. When a rotation parameter $a_{i}$ appears in a sum that runs from 1 to $N+1$, 
$a_{N+1}$ is taken to be zero.

The outermost horizon for the Kerr-AdS solution corresponds to the largest positive root of the equation $V(r_{h}) =
2m$. The area of the horizon is then given by
\begin{equation}\label{HorizonArea}
   A_{h} = \omega_{d-1} \,r_{h}^{\,1+\eps}\,\prod_{i=1}^{N} \frac{r_{h}^{\,2}+a_{i}^{\,2}}{\Xi_{i}} ~.
\end{equation}
where $\omega_{d-1}$ is the area of the unit $d-1$ sphere. The surface gravity $\kappa$ is obtained in terms of $V(r)$
and its first derivative, evaluated at the horizon
\begin{equation}\label{SurfaceGravity}
  \kappa = \frac{1}{2}\,\left(1+ \frac{r_{h}^{\,2}}{\ell^{2}}\right)\,\frac{V'(r_{h})}{V(r_{h})} ~.
\end{equation}
The Hawking temperature associated with the horizon is related to the surface gravity by $T = (2\pi)^{-1}\kappa$.
Evaluating $\kappa$ using \eqref{KerrV} gives
\begin{equation}\label{HawkingTemp}
  T = \frac{r_{h}}{2\pi}\,\left(a+\frac{r_{h}^{\,2}}{\ell^{2}}\right)\,\sum_{i=1}^{N}\frac{1}{r_{h}^{\,2}+a_{i}^{\,2}}
     + \frac{r_{h}}{2\pi \ell^{2}} + \frac{\eps-2}{4\pi r_{h}}\,\left(1+\frac{r_{h}^{\,2}}{\ell^{2}}\right) ~.
\end{equation}
Finally, it is useful to define a set of angular velocities relative to a non-rotating frame at infinity. For each
rotation parameter $a_{i}$ we define the associated angular velocity as
\begin{equation}\label{AngularVelocity}
   \Omega_{i} = \frac{a_{i}}{r_{h}^{\,2}+a_{i}^{\,2}}\,\left(1+\frac{r_{h}^{\,2}}{\ell^{2}}\right) ~.
\end{equation}

\pagebreak

\end{document}